\documentclass[apj]{emulateapj}
\usepackage{float,epsfig}
\usepackage{pstricks}

\def\pn{\par\noindent}
\def\chandra{{\it Chandra}}
\def\xmm{XMM--{\it Newton}}
\def\cgs{erg cm$^{-2}$ s$^{-1}$} 
\def\gsimeq{\hbox{\raise0.5ex\hbox{$>\lower1.06ex\hbox{$\kern-1.07em{\sim}$}$}}} 
\def\lsimeq{\hbox{\raise0.5ex\hbox{$<\lower1.06ex\hbox{$\kern-1.07em{\sim}$}$}}} 
\def\spose#1{\hbox to 0pt{#1\hss}}
\def\simlt{\mathrel{\spose{\lower 3pt\hbox{$\mathchar"218$}}
     \raise 2.0pt\hbox{$\mathchar"13C$}}}
\def\lsimeq{\mathrel{\spose{\lower 3pt\hbox{$\mathchar"218$}}
     \raise 2.0pt\hbox{$\mathchar"13C$}}}
\def\ls{\mathrel{\spose{\lower 3pt\hbox{$\mathchar"218$}}
     \raise 2.0pt\hbox{$\mathchar"13C$}}}
\def\simgt{\mathrel{\spose{\lower 3pt\hbox{$\mathchar"218$}}
     \raise 2.0pt\hbox{$\mathchar"13E$}}}
\def\gsimeq{\mathrel{\spose{\lower 3pt\hbox{$\mathchar"218$}}
     \raise 2.0pt\hbox{$\mathchar"13E$}}}
\def\gs{\mathrel{\spose{\lower 3pt\hbox{$\mathchar"218$}}
     \raise 2.0pt\hbox{$\mathchar"13E$}}}

\slugcomment{to appear in the COSMOS ApJS Special Issue, 2007}

\shorttitle{Optical identifications of X--ray sources in COSMOS}
\shortauthors{Brusa et al.}

\begin{document}

%% This is the end of the preamble.  Indicate the beginning of the
%% paper itself with \begin{document}.


%\twocolumn[  

\title{The XMM--{\it Newton} wide-field survey in the COSMOS field. III: \\ 
   optical identification and multiwavelength properties of a large sample of
   X--ray selected sources}\altaffiltext{$\star$}{Based on observations
   obtained with XMM-Newton, an ESA science mission with 
instruments and contributions directly funded by ESA Member States and NASA;
also based on data collected at: the NASA/ESA {\em
Hubble Space Telescope}, obtained at the Space Telescope Science
Institute, which is operated by AURA Inc, under NASA contract NAS
5-26555; the Subaru Telescope, which is operated by
   the National Astronomical 
Observatory of Japan; the European Southern Observatory, Chile, under Large
   Program 175.A-0839; Kitt Peak
   National Observatory, Cerro Tololo Inter-American Observatory, and the 
National Optical Astronomy Observatory, which are operated by the Association
   of Universities for Research in Astronomy, Inc. (AURA) under cooperative
   agreement 
with the National Science Foundation; and the Canada-France-Hawaii Telescope
   operated by the National Research Council of Canada, 
the Centre National de la Recherche Scientifique de France and the University
   of Hawaii.}


\author{M. Brusa\altaffilmark{1},
G. Zamorani\altaffilmark{2},
A. Comastri\altaffilmark{2},
G. Hasinger\altaffilmark{1},
N. Cappelluti\altaffilmark{1},
F. Civano\altaffilmark{2},
A. Finoguenov\altaffilmark{1},
V. Mainieri\altaffilmark{1,18},
M. Salvato\altaffilmark{1,8},
C. Vignali\altaffilmark{2}, 
M. Elvis\altaffilmark{3},
F. Fiore\altaffilmark{4},
R. Gilli\altaffilmark{2},
C.D. Impey\altaffilmark{5},
S.J. Lilly\altaffilmark{6},
M. Mignoli\altaffilmark{2},
J. Silverman\altaffilmark{1},
J. Trump\altaffilmark{5},
C.M. Urry\altaffilmark{7},
R. Bender\altaffilmark{1},
P. Capak\altaffilmark{8},
J.P. Huchra\altaffilmark{3},
J.P. Kneib\altaffilmark{9}
A. Koekemoer\altaffilmark{10},
A. Leauthaud\altaffilmark{9},
I. Lehmann\altaffilmark{1},
R. Massey\altaffilmark{8}, 
I. Matute\altaffilmark{1,17},
P.J. McCarthy\altaffilmark{11},
H.J. McCracken\altaffilmark{12,13},
J. Rhodes\altaffilmark{8},
N.Z. Scoville\altaffilmark{8,14},
Y. Taniguchi\altaffilmark{15},
D. Thompson\altaffilmark{8,16}}

\altaffiltext{1}{Max Planck Institut f\"ur extraterrestrische Physik, 
       Giessenbachstrasse 1, D--85748 Garching, Germany} 
\altaffiltext{2}{INAF --  Osservatorio Astronomico di Bologna, via Ranzani 1, 
I--40127 Bologna, Italy}
\altaffiltext{3}{Harvard-Smithsonian Center for Astrophysics, 60 Garden
Street, Cambridge, MA 02138 } %elvis
\altaffiltext{4}{INAF --  Osservatorio Astronomico di Roma, via Frascati 33,
  Monteporzio-Catone (Roma), I-00040, Italy}
\altaffiltext{5}{Steward Observatory, University of Arizona, 933 North
Cherry Avenue, Tucson, AZ 85 721} %Impey
\altaffiltext{6}{Department of Physics, Eidgenossiche Technische Hochschule
  (ETH), CH-8093 Zurich, Switzerland}
\altaffiltext{7}{Yale Center for Astronomy and Astrophysics, Yale University,
  P.O. Box 208121, New Haven CT 06520-8121, USA}
\altaffiltext{8}{California Institute of Technology, MC 105-24, 1200 East
California Boulevard, Pasadena, CA 91125} %Scoville, Capak, Thompson
\altaffiltext{9}{Laboratoire d'Astrophysique de Marseille, BP 8, Traverse
du Siphon, 13376 Marseille Cedex 12, France}
\altaffiltext{10}{Space Telescope Science Institute, 3700 SanMartin
Drive, Baltimore, MD 21218} % Koekemoer
\altaffiltext{11}{Carnegie Observatories, 813 Santa Barbara Street, Pasadena,
  CA 91101}
\altaffiltext{12}{Institut d'Astrophysique de Paris, UMR7095 CNRS,
  Universit\'e Pierre \& Marie Curie, 98 bis boulevard Arago, 75014 Paris,
  France} 
\altaffiltext{13}{Observatoire de Paris, LERMA, 61 Avenue de l'Observatoire,
  75014 Paris, France} 
\altaffiltext{14}{Visiting Astronomer, Univ. Hawaii, 2680 Woodlawn Dr.,
Honolulu, HI, 96822} %Scoville
\altaffiltext{15}{Physics Department, Graduate School of Science and
         Engineering, Ehime University, 2-5 Bunkyo-cho, Matsuyama, Ehime
         790-8577. Japan}
\altaffiltext{16}{Large Binocular Telescope Observatory
   University of Arizona, 933 N. Cherry Ave.
   Tucson, AZ  85721-0065,   USA} % Thompson
\altaffiltext{17}{INAF -- Osservatorio Astrofisico di Arcetri, Largo E. Fermi,
  5, 50125 Firenze, Italy} %Israel
\altaffiltext{18}{European Southern Observatory, Karl-Schwarzschild-str. 2,
  85748 Garching bei M\"unchen, Germany}


\begin{abstract}
We present the optical identification of a sample of 695 
X-ray sources detected in the first 1.3 deg$^2$ of the XMM-COSMOS survey, down
to a 0.5-2 keV (2-10 keV) limiting flux of $\sim10^{-15}$ erg cm$^{-2}$ s$^{-1}$  
($\sim5\times10^{-15}$ erg cm$^{-2}$ s$^{-1}$).  
In order to identify the correct optical
counterparts and to assess the statistical significance of the
X-ray to optical associations we have used 
the  ``likelihood ratio technique''.
Here we present the identification method and its application to the
CFHT I-band and photometric catalogs. We were able to  associate a 
candidate optical counterpart to $\sim$90\% (626) of the X-ray 
sources, while for the remaining $\sim$10\% of the sources we were not able
to provide a unique optical association due to the faintness of the
possible optical counterparts (I$_{\rm AB}>$25) or to the presence of multiple
optical sources, with similar likelihoods of being the correct identification,
within the XMM--{\it Newton} error circles. % (about 3-5$"$).
We also cross-correlated the  candidate optical counterparts with the
Subaru multicolor and ACS catalogs and with the Magellan/IMACS, zCOSMOS and
literature spectroscopic data; the spectroscopic sample comprises 248 objects
($\sim$40\% of the full sample). 
Our analysis of this statistically meaningful sample of X--ray sources reveals
that for $\sim$80\% of the counterparts there is a very
good agreement between the spectroscopic classification, the morphological
parameters as derived from ACS data, and the optical to near infrared
colors: the large majority of spectroscopically identified broad line AGN (BL
AGN) have
a point-like morphology on ACS data, blue optical colors in color-color
diagrams, and an X--ray to optical flux ratio typical of optically selected
quasars. Conversely, sources classified as narrow line AGN or normal galaxies
are on average associated with extended optical sources, have significantly
redder optical to near infrared colors and span a larger range of X--ray to
optical flux ratios.  
%{\bf The majority ($\gsimeq 80$\%) of X-ray selected point-like quasars
%occupy the QSO locus in classic optical color-color plot 
%and would have been selected as outliers from the stellar locus.}
However, about 20\% of the sources show an apparent mismatch between the morphological
and spectroscopic classifications. All the ``extended'' BL AGN lie at redshift $<$1.5, while the redshift
distribution of the full BL AGN population peaks at z$\sim$1.5. 
The most likely explanation is that in these objects the nuclear emission is not dominant  with
respect to the host galaxy emission in the observed ACS band. 
Our analysis also suggests that the Type 2/Type 1 ratio decreases towards 
high luminosities, in qualitative agreement with the results from X--ray
spectral analysis and the most recent modeling of the X--ray luminosity
function evolution.
%Our analysis indicates that the observed
%differences are largely explained by the location of these objects in the
%redshift--luminosity plane. 
\end{abstract}



\section{Introduction}
\pn
The primary goals of the ``Cosmic Evolution
Survey'' (COSMOS, \citealt{scoville1}) is to trace star formation and nuclear
activity along  with the mass assembly history of galaxies as a function of
redshift and environment. 
Although there were early theoretical suggestions (e.g. \citealt{sr1998}), it
was  
the tight relation observed in local galaxies between the
black holes mass and the velocity dispersion (the M$_{\rm BH}$-$\sigma$  
relation, see e.g. \citealt{f2000,g00}) 
and the fact that the locally inferred black hole mass density
appears to be broadly consistent with the estimates of the mass accreted
during the quasar phase \citep{fi99,yt02,erz,marconi,merloni} that made it
clear that BH--driven nuclear activity  
and the assembly of bulge masses are closely linked.
This realization has led to a large number of
theoretical studies, suggesting feedback mechanisms to explain this
fundamental link between the assembly of black holes (BH) and the formation of
spheroids in galaxy halos (\citealt{dimatteo,menci} and reference therein). 

\pn
In this framework, the hard X--ray band is by far the 
cleanest one for studying the history of accretion onto black holes in the
Universe, being 
the only band in which emission from accretion processes clearly dominates the 
cosmic background. 
The detailed study of the nature of the hard X--ray source population
is being pursued by complementing deep pencil beam observations 
with shallower, larger area surveys (see \citealt{bh05} for a recent review). 
Indeed, in the recent years, large efforts have been
dedicated to the optical to radio characterization of X--ray sources selected
at different X--ray depths (see, among the most recent: the XBo\"otes survey
\citep{brand}, the Serendipitous Extragalactic X--ray Source Identification
SEXSI \citep{sexsi}, the Extended Groth strip EGS \citep{georgakakis}, 
the HELLAS2XMM survey \citep{cocchia}).
In particular, sizable samples of objects detected at the bright X--ray fluxes
($\gtrsim 10^{-14}$ \cgs) over an area of the order of a few  square degrees
are needed to cover the high luminosity part of the Hubble diagram and
to obtain, together with samples from narrower and deeper pencil--beam
observations, a well constrained luminosity function with a similar number 
of sources per luminosity decade and redshift bin.
The results from both deep \citep{alex03,cdfn,g02,cdfs,lockman} and shallow
surveys \citep{f03,green04,clasxs} have unambiguously unveiled a differential
evolution for the low-- and high--luminosity AGN population
\citep{ueda,clasxs2,silverman,has05,lafranca}.  
However, in these studies, the evolution of the  high-luminosity tail of
the obscured AGN luminosity function remains a key parameter still to be
determined.  
The best strategy to address this issue is to increase 
the area covered in the hard X--ray band, and the corresponding 
optical-NIR photometric and spectroscopic follow-up, down to 
F$_{2-10keV}\sim5\times10^{-15}$ \cgs, where the bulk of the 
XRB is produced. \\
Another important issue in AGN studies is the determination of the global
Spectral Energy Distributions (SEDs), over the widest possible frequency
range, of different types of AGN.  Large samples of AGN have been assembled
using different selection criteria at different frequencies (e.g. radio,
infrared, optical-UV, X-ray); however, the lack of complete multi-wavelength
coverage for many of these samples makes it difficult the comparison of the
properties of different classes of AGN. For example, while the average 
SED of optically and radio selected bright quasars has been reasonably well known
for more than a decade \citep{elvis94}, little is still known about the
SEDs of lower luminosity X-ray selected AGNs and, in particular, of the obscured
ones. This reflects into a significant uncertainty in the estimate of  the
bolometric correction  ($k_{bol}$, i.e. the correction from the X--ray to the
bolometric luminosity) to be applied to the observed luminosity
of  X-ray selected AGN and, in turn, on the estimates of the Eddington
luminosities and of the masses of the central black holes \citep{fabian03}. 
Recent results on high--redshift quasars 
have shown that the overall X--ray to optical spectral slope 
($\alpha_{\rm ox}$, usually measured between 2500 \AA\ and 2 keV, e.g. 
\citealt{tananbaum}) and the corresponding $k_{bol}$, is a function of the  
AGN luminosity, with the low luminosity objects having a lower $k_{bol}$
\citep{vignali,fabian03,steffen06}. It is therefore clear that, in order to
get a complete census of the luminosity output of AGN, a multi-wavelength
project is needed: only in this way can the SEDs from
radio to X-rays of a large sample of AGN, selected at different frequencies,
be compiled. \\ 
\pn
The  XMM--{\it Newton} wide-field survey in the COSMOS field (hereinafter:
XMM-COSMOS) is an important step forward in addressing the
topics described above.
The $\sim2$ deg$^2$ area of the HST/ACS COSMOS Treasury program \citep{scoville2}, bounded by
$9^{\rm h}57.5^{\rm m}<$R.A.$<10^{\rm h}03.5^{\rm m}$; $1^{\rm d}27.5^{\rm
  m}<$DEC$<2^{\rm d}57.5^{\rm m}$, 
has been surveyed with \xmm\ for a total of $\sim$800 ks during AO3, and
additional 600 ks have been already granted in AO4 to this project. 
When completed, XMM-COSMOS will provide an unprecedently large sample of X-ray
sources ($\gsimeq2000$), detected on a large, contiguous area, with {\it
  complete} ultraviolet to mid-infrared (including Spitzer data) and radio coverage, and
almost complete spectroscopic follow--up granted through the zCOSMOS
\citep{lilly} and Magellan/IMACS \citep{impey} projects. \\
The XMM-COSMOS project is described in \citet{papI}.
The X--ray point source counts from the first 800 ks of the \xmm\ observations obtained so
far are presented in \citet{papII}, while the properties
of extended and diffuse sources are described in \citet{alexis}.
First results from the spectral analysis of AGN with known redshift and high
counting statistics are presented in \citet{papIV}. \\
This paper presents the optical identification of X--ray point sources detected in
this first year of the \xmm\ data, and the first results on the multiwavelength
properties of this large sample of X-ray selected AGN.\\ 
The paper is organized as follows: Section~2 presents the multiwavelength
datasets and  describes the method used to identify
the X-ray sources and its statistical reliability; the X--ray to optical and
near infrared properties are presented in Sect.~3, along with preliminary
results from the ongoing spectroscopic follow--up; in Sections 4 and 5 
we discuss and summarize the most important results. 
Throughout the paper, we adopt the cosmological parameters $H_0=70$ km s$^{-1}$
Mpc$^{-1}$, $\Omega_m$=0.3 and $\Omega_{\Lambda}$=0.7 \citep{spergel}.

%%%%%%%%%%%%%%%%%%%%%%%%%%%%%%%%%%%%%%%%%%%%%%%%
%\begin{center}
\begin{figure*}
\includegraphics{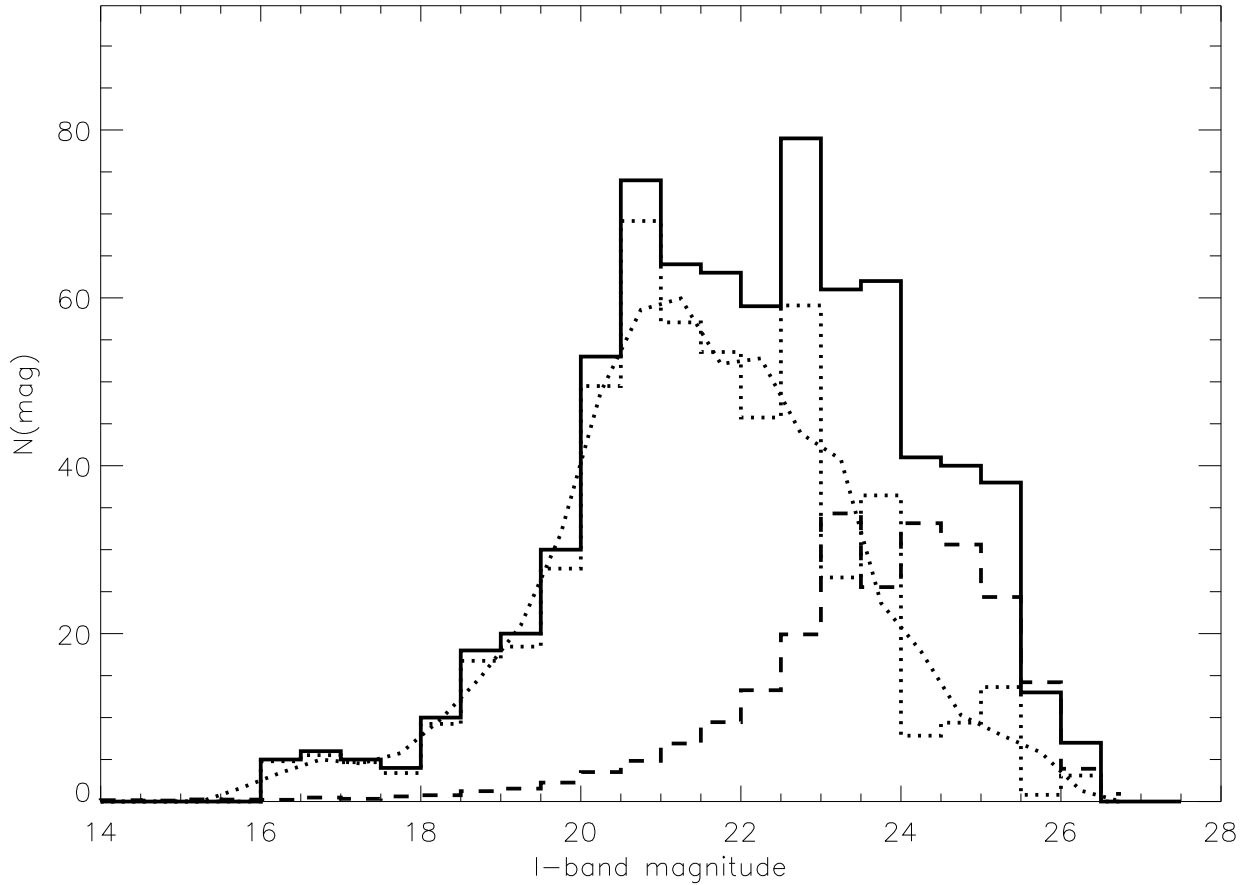}
\caption{Observed magnitude distribution of all the
optical objects detected in the $I$ band within a radius of 3\arcsec\
around each of the 674 point-like X--ray sources fainter than I$_{\rm AB}$=16 
(solid histogram), together with the expected distribution of background 
objects in the same area ({\it n(m)}, dashed histogram). 
The difference between these two
distributions (dotted histogram) is the expected magnitude
distribution of the optical counterparts. The smooth curve fitted to
this histogram  (dotted line) has been used as input in
the likelihood calculation ({\it q(m)}).}
\label{histo}
\end{figure*}
%\end{center}
%%%%%%%%%%%%%%%%%%%%%%%%%%%%%%%%%%%%%%%%%%%%%%%%

\section{X-ray source identification}

\subsection{Optical and X--ray  datasets}
\pn
The XMM-COSMOS X-ray source catalog comprises 1390 
different point--like X--ray sources detected over an area of $\sim$2 deg$^2$
in the first-year  \xmm\ observations of the COSMOS field \citep{papI,papII}. 
In this paper, we limit our analysis to the sources detected in the first 12 XMM-COSMOS
observations (over a total of $\sim$1.3 deg$^2$), for which both the X-ray
source list and the optical catalogs were in place in June 2005, and a
substantial spectroscopic follow-up already exists. These fields are flagged
in Table 1 in \cite{papI}, and the average exposure time is of $\sim 25 ks$
($\sim 40$ ks in the overlapping region, see \citealt{papI} for more details).
The catalog paper with the identification from the entire XMM-COSMOS survey is
in preparation and will rely strongly on the identification procedure described here.\\
The 12 fields X-ray catalog 
(hereinafter: 12F catalog) comprises 715 X-ray sources 
detected over a total of $\sim$1.3 deg$^2$.  
From the 12F catalog we removed 20 sources 
classified as extended by the detection algorithm (see \citealt{papII} for
a description of the adopted detection algorithm). The observed X--ray 
emission from most of these sources is likely to be due either to
groups/clusters of galaxies, or to the contribution of two or more X--ray
sources close to each other. In both cases, an association 
with a unique optical counterpart is not possible. The total number of
point-like X-ray sources in the 12F catalog, detected in at least one of the 
X-ray bands is therefore 695. Of these, 
656 are detected in the soft band (0.5-2 keV), 312 in the medium band (2-4.5
keV; 38 only in the medium band), and 47 in the hard band (4.5-10 keV; 1 only
in the hard).
The X--ray centroids of these 695 point-like X-ray sources 
have been astrometrically calibrated using the SAS task {\tt eposcorr},
as described in \citet{papII}; 
the resulting shift of $\sim$1$''$ ($\Delta$($\alpha$)=0.99$''$;
$\Delta$($\delta$)=0$''$) 
was applied to all source positions.\\
As a first step in the identification process, we used the I--band
CFHT/Megacam catalog \citep{megacam}, which, although  slightly shallower
than other available data  (e.g. the Subaru B,g,V,R,I and z photometric data,
see \citealt{capak}), has the advantage of having reliable photometry even at
bright magnitudes (I$_{\rm AB}\lsimeq19$), where the Subaru photometry starts
to be significantly affected by uncertainties due to saturation.\\
At magnitudes brighter than I$_{\rm AB}=16$, only $\sim$0.24 sources 
are expected by chance in a 3$''$ error-circle around all the 695 
X-ray sources on the basis of the background counts of the Megacam catalog;
therefore we considered as secure optical identifications all 
the 21 sources brighter than I$_{AB}$=16 in the CFHT catalog and within 3$''$
from the X--ray centroids. % 40  
For all the remaining X--ray sources (674) %1293 
we used the method described in Sect.~\ref{xrid}. \\
Then, we have cross-correlated the optical counterparts with the multicolor
photo-z catalog (June 22th 2005 release, \citealt{capak})\footnote{The new
  version of the photometric catalog has been released at the time of the
  submission of this paper. The (small) differences between the two
  photometric catalogs do not affect the results presented in this work.}.
All the magnitudes are in the AB system \citep{oke}, if not otherwise
stated.\footnote{We adopted the following AB-Vega conversion:
    I(Vega)=I(AB)-0.4, 
R-K(Vega)=R-K(AB)+1.65.}

\subsection{The method}
\label{xrid} 
\pn
Typical error-circles of XMM--{\it Newton} data ($\sim 5''$ arcsec radius at
95\% confidence level\footnote{Such a radius represents the radius for which
  $\sim$95\% of the XMM-Newton sources in the SSC catalog are associated with
  USNO A.2 sources (see Fig. 7.5 in \citealt{firstxmm})}) often contain more 
than one source in deep optical images, so that the identification process is
not always straightforward. Obviously, this is particularly true when the
candidate counterparts are faint.
We therefore decided to use the ``likelihood ratio'' 
($LR$) technique, in order to properly identify the optical counterparts
\citep{ss92,ciliegi2003,brusa05}.
The $LR$ is  defined as the ratio between the probability that the source is
the correct identification and the corresponding probability of being a
background, unrelated object \citep{ss92}, i.e.: \\
\centerline{$LR = \frac{q(m) f(r)}{n(m)}$}\\
where {\it q(m)} is the expected probability distribution, as a
function of magnitude, of the true counterparts, {\it f(r)} is the probability
distribution function of the positional errors of the X--ray sources assumed
to be a two--dimensional Gaussian, and {\it n(m)} is the surface density of
background objects with magnitude {\it m}. 
For the calculation of the $LR$ parameters 
we followed and improved the procedure described in \citet{brusa05}. 
For the {\it f(r)} calculation, we used the statistical error as computed
  from the detection procedure and tested against the pattern of observations 
   with extensive Monte Carlo simulations presented by \citet{papII}. 
  We also added in quadrature a systematic error
  of 0.75$''$: as noticed by Loaring et al. (2005), this additional component may be
  due to residual uncertainties in the detector geometry, and may represent a
  fundamental limit to the positional accuracy of \xmm.
We adopted a 3$''$ radius for the estimate of {\it q(m)}, obtained by subtracting the
expected number of background objects ({\it n(m)}) from the observed total number of
objects listed in the catalog around the positions of the X-ray sources. 
Since on the basis of several results in the literature (see
e.g. \citealt{f03,dellaceca,l05}), a large fraction of the possible counterparts are
expected to be found within a 3$''$ radius, this choice maximizes the
statistical significance of the over-density around the X-ray centroids, due
to the presence of the optical counterparts.  
With this procedure, {\it q(m)} is well defined up to I$_{\rm
  AB}\sim$23.0-23.5. At fainter magnitudes, the number of
objects in the error boxes around the X-ray sources turns out to be smaller than that expected from
the field global counts {\it n(m)}. Formally, this would produce an unphysical
negative {\it q(m)}, which, in turn, would not allow the application of this
procedure at these magnitudes. The reason for this effect is the presence of a
large  number of relatively bright optical counterparts (I$_{\rm AB}$ =
16-23) close to the X-ray centroids. These objects, which occupy a non-negligible
fraction of the total area of the X-ray error boxes, make it difficult to
detect fainter background objects in the same area. As a consequence, the {\it
  n(m)} estimated from the global field is an overestimate of the observed {\it
  n(m)} at faint magnitudes in the X-ray error boxes. In order
to estimate the correct {\it n(m)} to be used at faint magnitudes in the
likelihood calculation, we have randomly  
extracted from the Megacam catalog 1000 optical sources
with the same expected magnitude distribution of the X-ray sources, and we 
computed the background surface density around these objects. 
As expected, we found that, indeed, the {\it n(m)} computed in this way  is
consistent with the global {\it n(m)} at I$_{\rm AB}$$<$23.0, but is
significantly smaller than it (and smaller than the observed counts in the
error boxes) at fainter magnitudes. Therefore, the input {\it n(m)} 
in the likelihood procedure was the global one for I$_{\rm AB}$$<$23 and that
derived 
with this analysis around sources with the same magnitude distribution as the
``bright'' optical counterparts for I$_{\rm AB}>$23.0. This allowed us to
identify a few tens of very faint sources that would have been missed 
without this correction in the expected {\it n(m)}. \\

%%%%%%%%%%%%%%%%%%%%%%%%%%%%%%%%%%%%%%%%%%%%%%%%
%\begin{center}
\begin{figure}
%\begin{center}
\includegraphics[width=2.5cm]{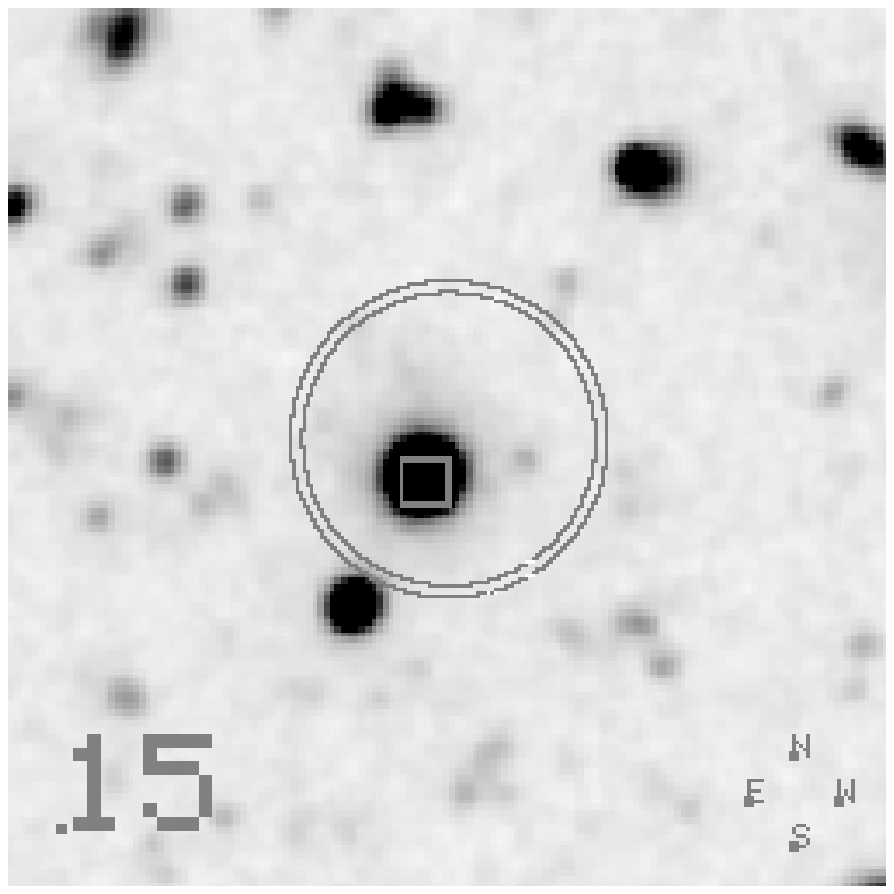}
\includegraphics[width=2.5cm]{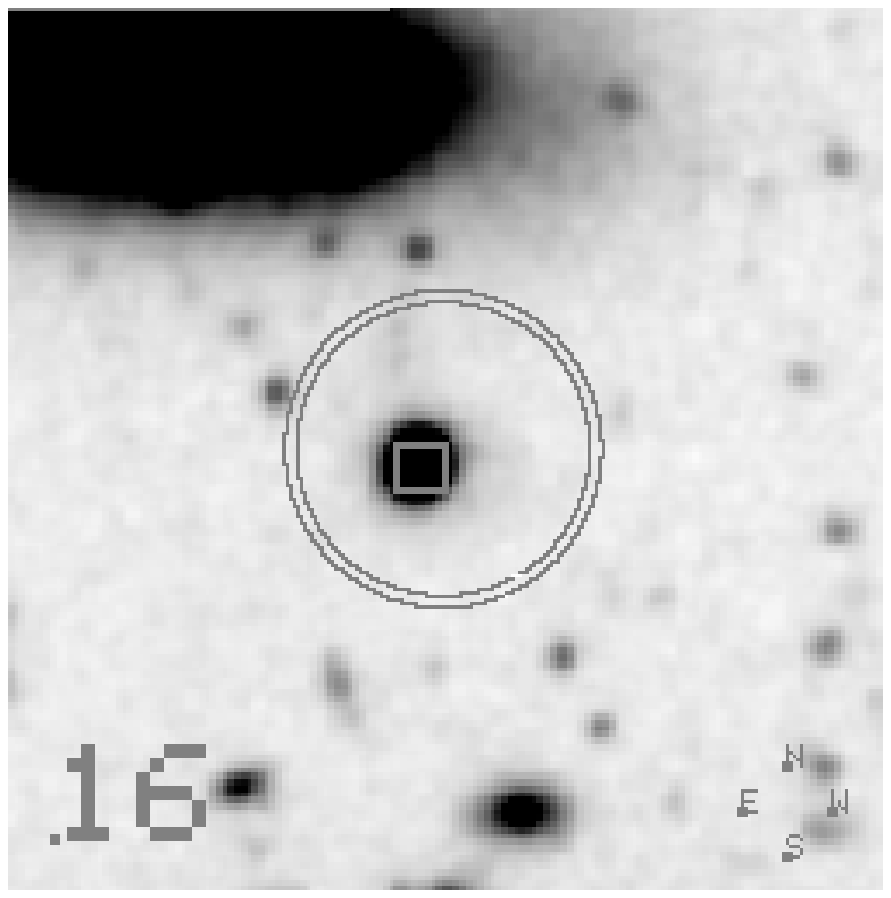}
\includegraphics[width=2.5cm]{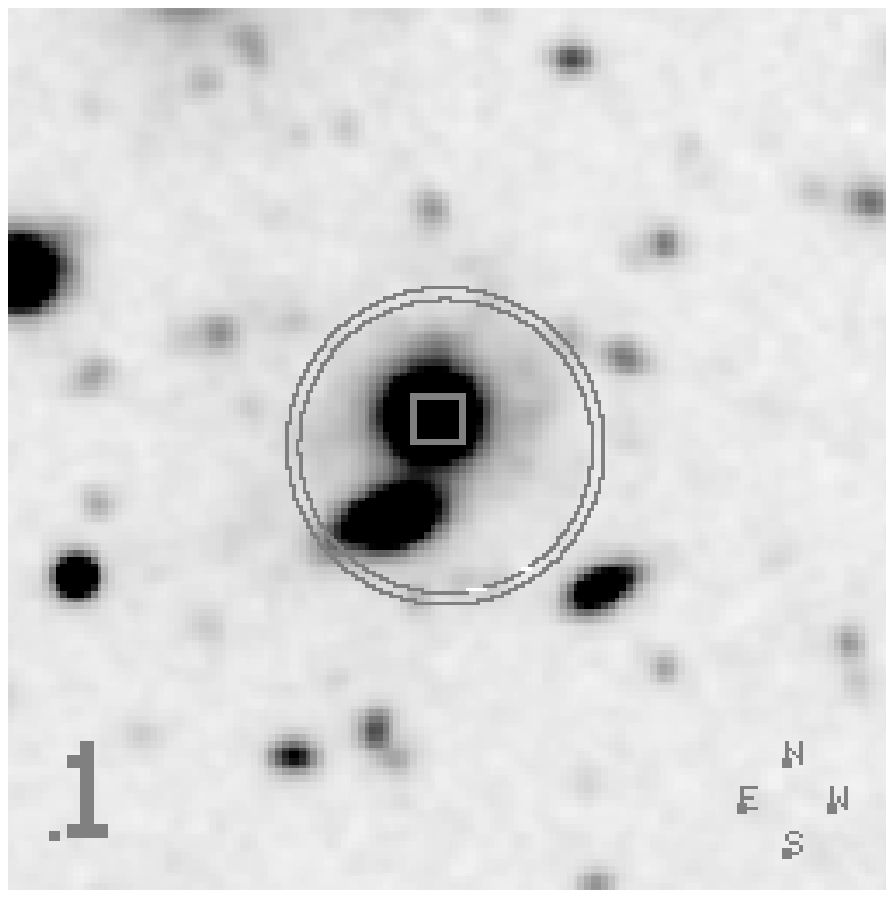}
\includegraphics[width=2.5cm]{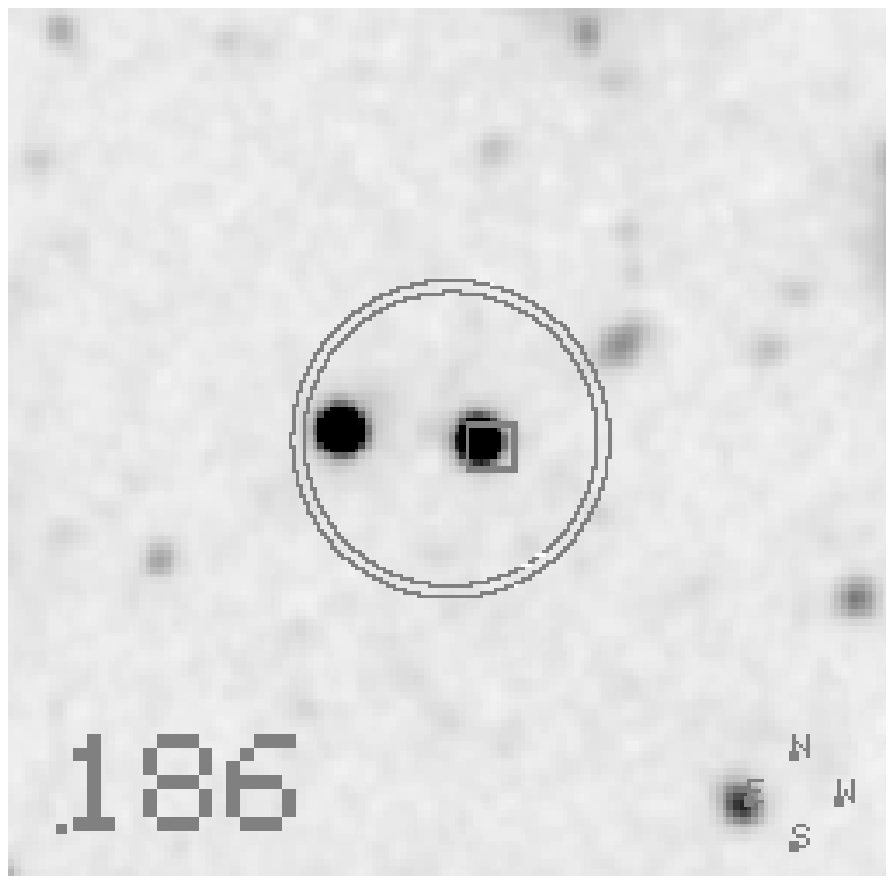}
\includegraphics[width=2.5cm]{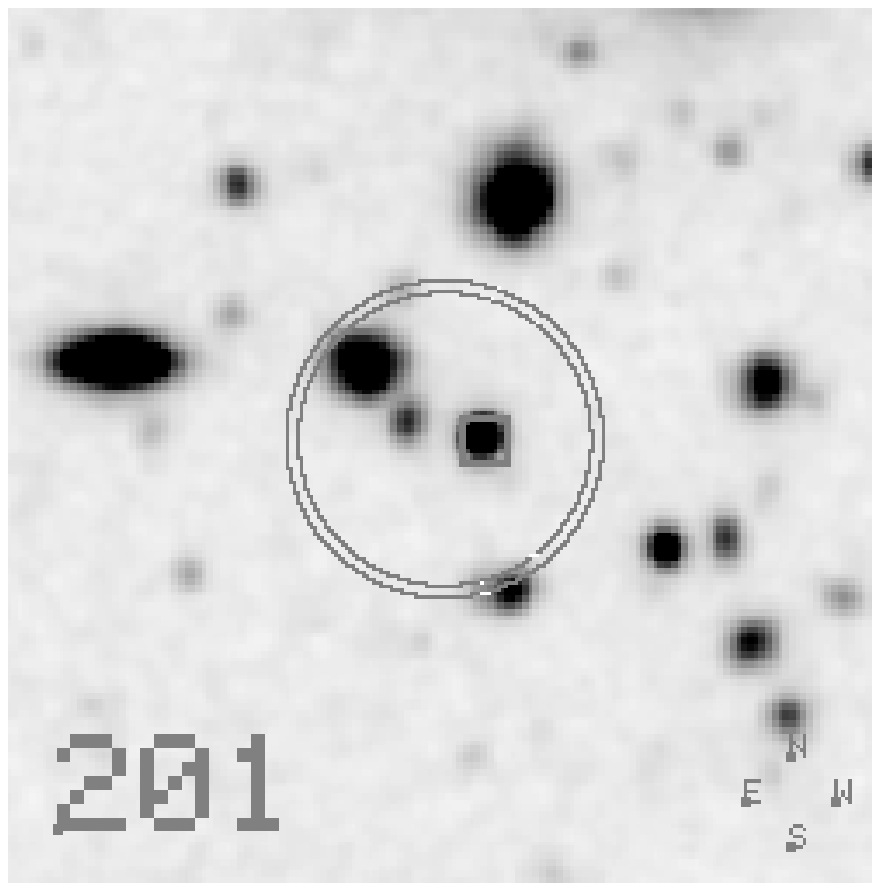}
\includegraphics[width=2.5cm]{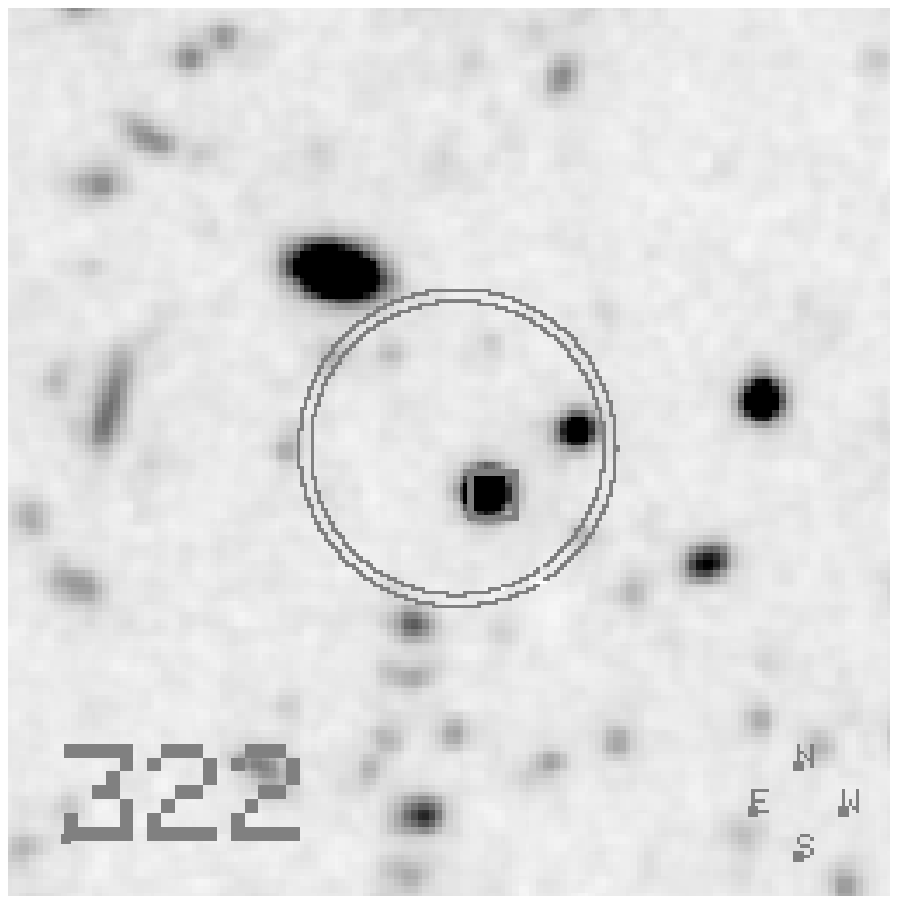}
\includegraphics[width=2.5cm]{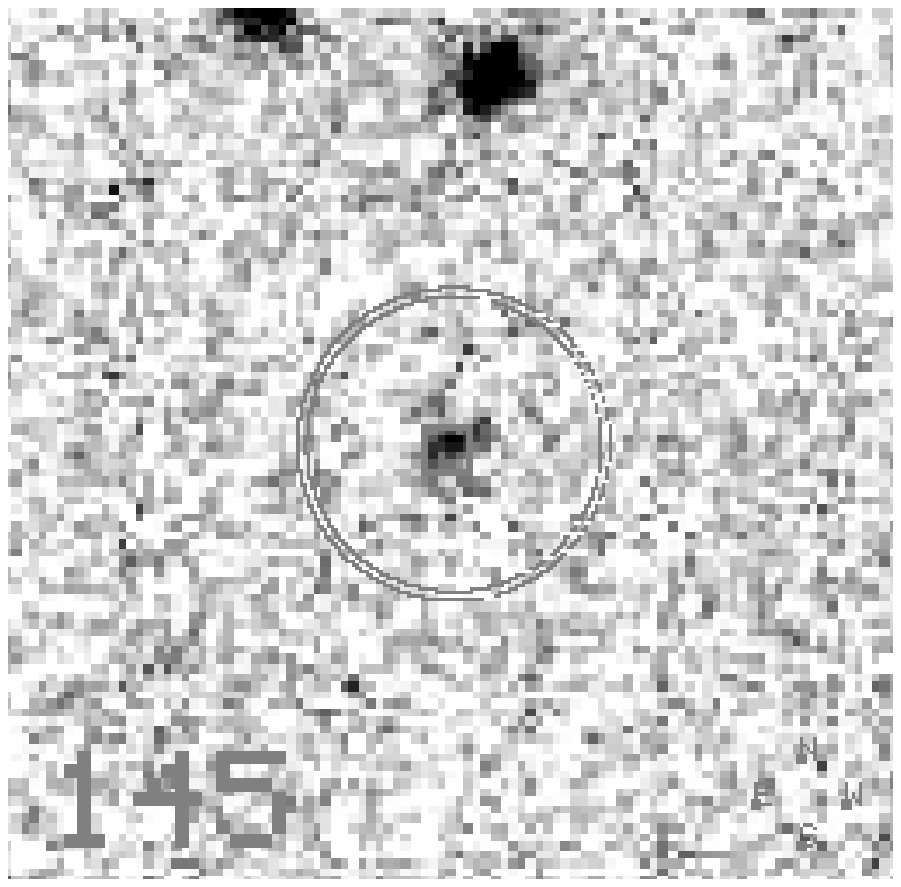}
\includegraphics[width=2.5cm]{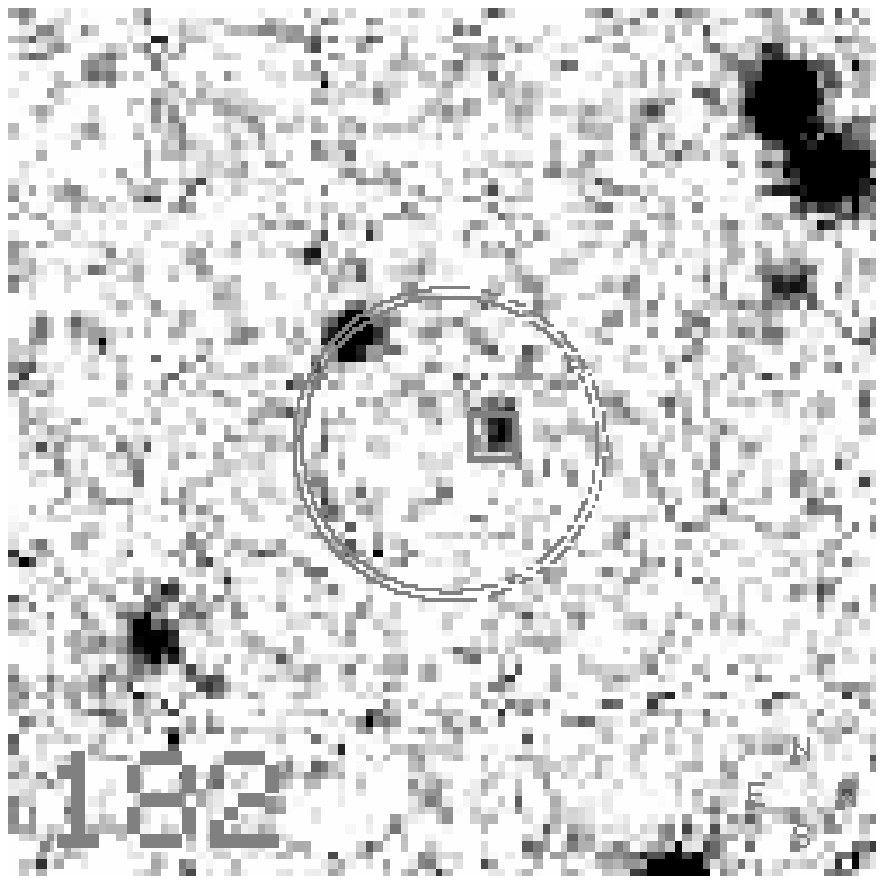}
\includegraphics[width=2.5cm]{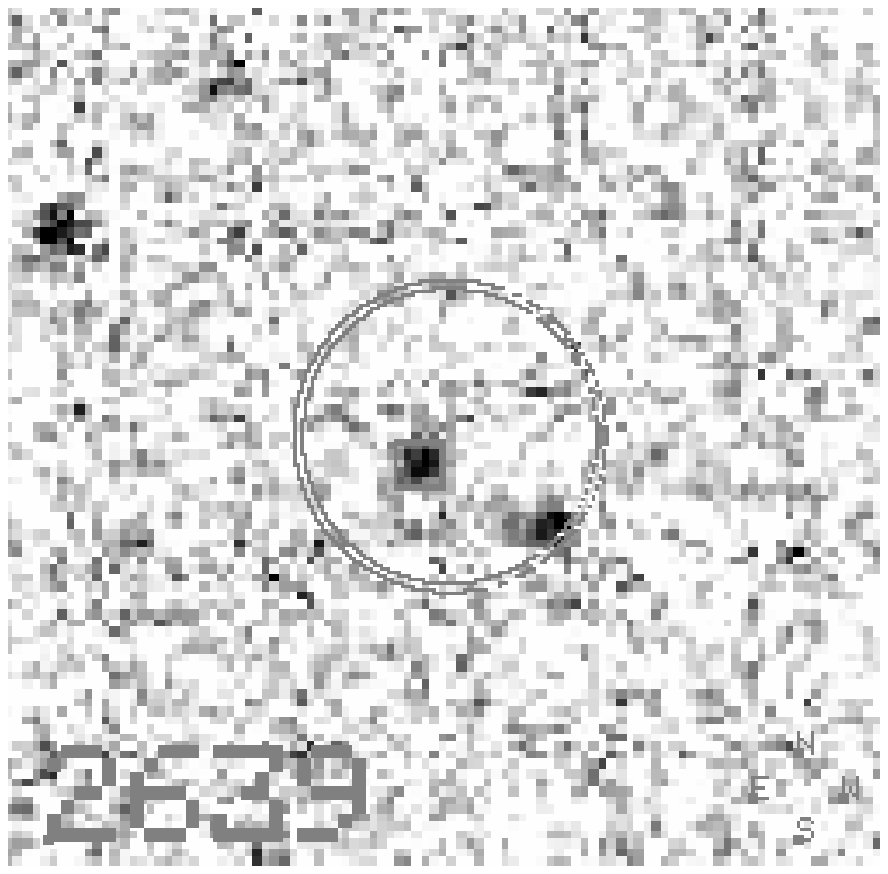}
%\end{center}
\caption{{\it (Upper panel)}: examples of I--band finding charts of sources
with a unique optical counterpart or securely identified from the likelihood
analysis. {\it (Middle panel)}: examples of I--band finding charts of sources with ambiguous 
counterparts. {\it (Lower panel)}: examples of K--band finding charts of sources 
identified from the K-band catalog. In all the panels, the circle is centered
on the \xmm\ position and has a radius of 5$''$.}
\label{fc}
\end{figure}
%\end{center}

\begin{deluxetable*}{lccccc}
\tablecaption{Summary of optical identification of point-like X--ray sources}
\tablehead{
\colhead{} & \colhead{all} & \multicolumn{2}{c}{identified} & \colhead{unidentified}} 
\startdata
Sources with I$<16$ (I bright) & 21 & \multicolumn{2}{c}{21} & - \\
Sources with I$>16$ & 674 &  \multicolumn{2}{c}{587} & 87\\
Additional sources identified with K & - &  \multicolumn{2}{c}{46} & - \\
\hline
Total & 695 &  \multicolumn{2}{c}{654}  &  41 \\
 &  & reliable\tablenotemark{a} & ambiguous\tablenotemark{b} &  \\
 &  & 626 & 28 &  \\
\enddata
\tablenotetext{a}{We classified as ``reliable'' all the 21 I-bright sources, the
46 sources identified from the K-band, and 559 objects for which there is only
one object with LR$>LR_{\rm th}$ or the ratio between the highest and
the second highest LR value in the I and/or in the K-band is greater than
3. See Sect.~2.3 for details.}
\tablenotetext{b}{We classified as ``ambiguous'' all the sources for which the
  ratio of LR values of the possible counterparts in the
  I-band {\it and} in the K-band is smaller than 3. See Sect.~2.3 for
  details.}  
\end{deluxetable*}

\pn
Fig.~1 shows the observed magnitude distribution of all optical objects 
present in the $I$ band catalog within a radius of 3\arcsec\
around each X--ray source (solid histogram), together with the
expected distribution of background objects in the same area, estimated using 
the procedure described above (dashed histogram). 
The difference between these two distributions
(dotted histogram) is the expected magnitude distribution
of the optical counterparts. The smooth curve fitted to this histogram
(dotted curve) has been used as the input in the likelihood
calculation {\it q(m)}; the normalization of this curve has been set to 0.76,
corresponding to the ratio between the integral of the {\it q(m)} distribution
and the total number of the X--ray sources. For the threshold value in the
likelihood ratio we adopted $LR_{\rm th}$=0.4, which maximizes the sum of
sample reliability and completeness in the present sample \citep{ss92,brusa05}.\\

%%%%%%%%%%%%%%%%%%%%%%%%%%%%%%%%%%%%%%%%%%%%%%%%
%\begin{center}
\begin{figure}
\begin{center}
\includegraphics[width=7.5cm]{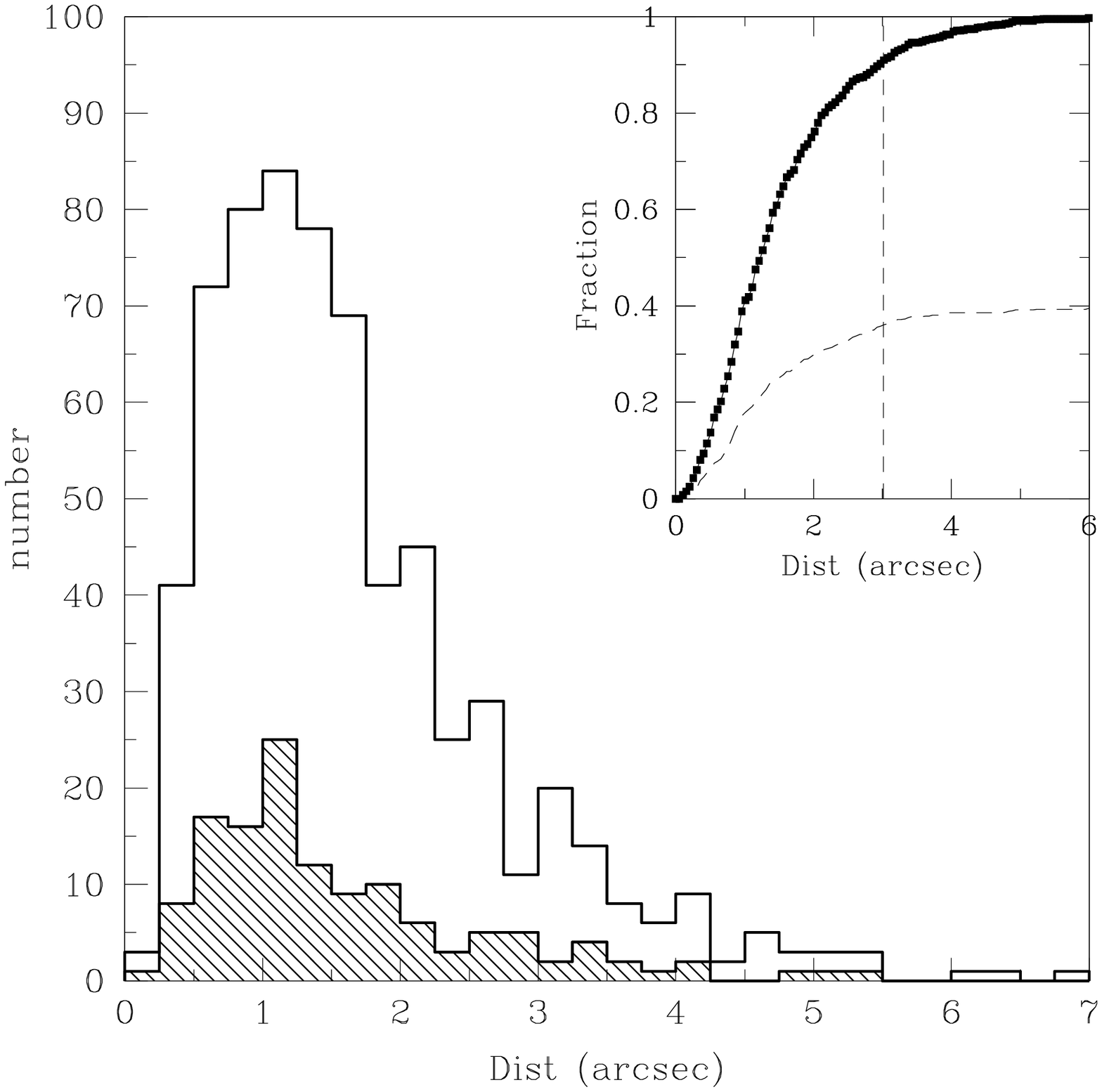}\\
\includegraphics[width=7.5cm]{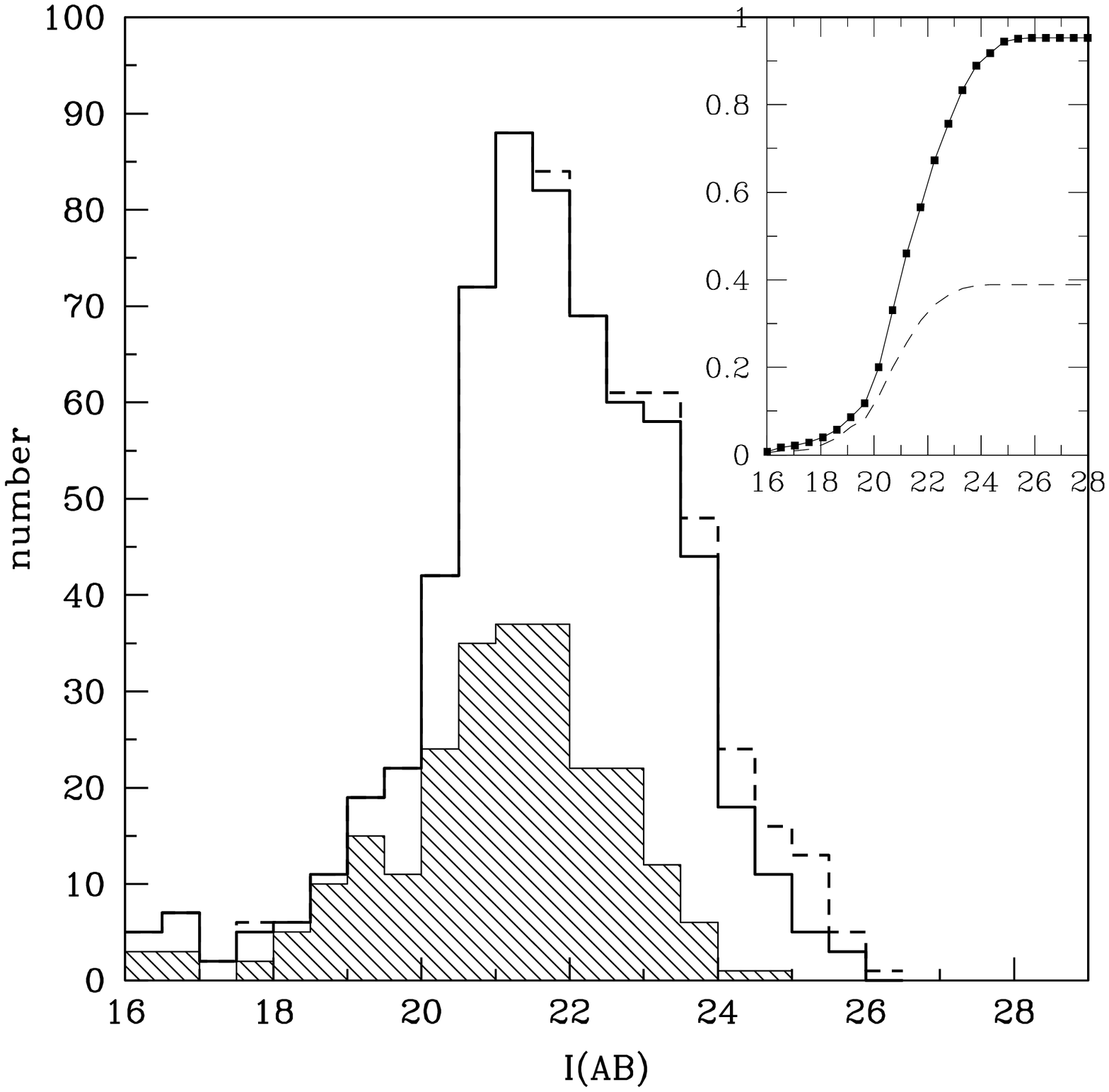}
\end{center}
\caption{
({\it Upper panel}): Histogram of the distances between the X--ray
  and optical counterparts for all the 654 sources identified as discussed in
  Sect. 2.3 (solid histogram) and the subsample of sources with spectroscopic
  redshift and classification (shaded histogram). The inset in the top-right
  part of the plot shows the cumulative distribution of X--ray to optical
  distances for all the sources (bold curve) and for the subsample of sources
  with spectroscopic redshift (dashed curve). The dashed line at 3$''$ marks
  the radius which encloses 90\% of the X--ray counterparts (see discussion in
  Section 2.2). 
  ({\it Lower panel}): Histogram of the I-band 
  magnitudes distribution for the 633 primary identifications fainter than
  I$_{\rm AB}$=16. The solid and
  shaded histograms have the same meaning as in the upper panel. The dashed
  histogram shows the lower limit to the I$_{\rm AB}$ magnitudes,
  corresponding to the brightest objects in the error-circles, for the 41
  X--ray sources without an optical identification. 
  The inset in the top-right part of the plot shows the cumulative
  distribution of X--ray to optical distances for all the sources with I>16 
  (bold curve) and for the subsample of sources with spectroscopic redshift
  (dashed curve).   }  
\label{maghisto}
\end{figure}
%\end{center}
%%%%%%%%%%%%%%%%%%%%%%%%%%%%%%%%%%%%%%%%%%%%%%%%

%%%%%%%%%%%%%%%%%%%%%%%%%%%%%%%%%%%%%%%%%%%%%%%%
%\begin{center}
\begin{figure*}
\centerline{\includegraphics[width=16cm]{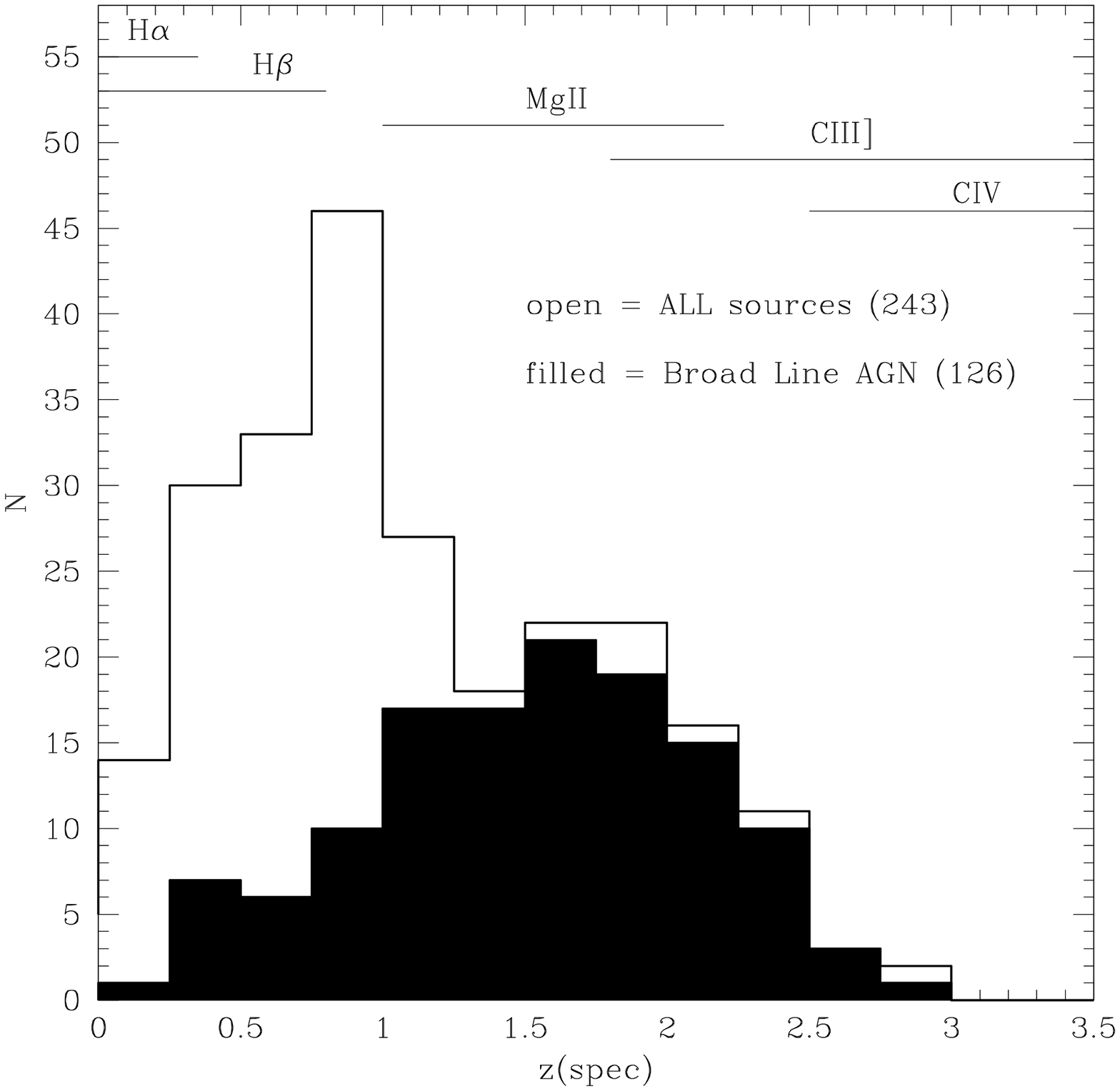}}
\caption{Redshift distribution for the sample of 243 sources with
  ``secure'' spectroscopic information from Magellan/IMACS data 
  \citep{trump}), zCOSMOS project \citep{lilly} and from the SDSS archive
  \citep{sdss}. The black filled histogram shows the contribution of
  spectroscopically classified  Broad Line AGN. 
  The redshift windows in which each broad line (H$\alpha$, H$\beta$,
  MgII, CIII], CIV) is visible in the IMACS/zCOSMOS spectra
  (wavelength range $\sim$5400-9000 \AA) are also reported 
  in the upper part of the Figure.} 
\label{zdist}
\end{figure*}
%\end{center}
%%%%%%%%%%%%%%%%%%%%%%%%%%%%%%%%%%%%%%%%%%%%%%%%%%%%%%%%%
%\begin{center}
\begin{figure}
\centerline{\includegraphics[width=9.5cm]{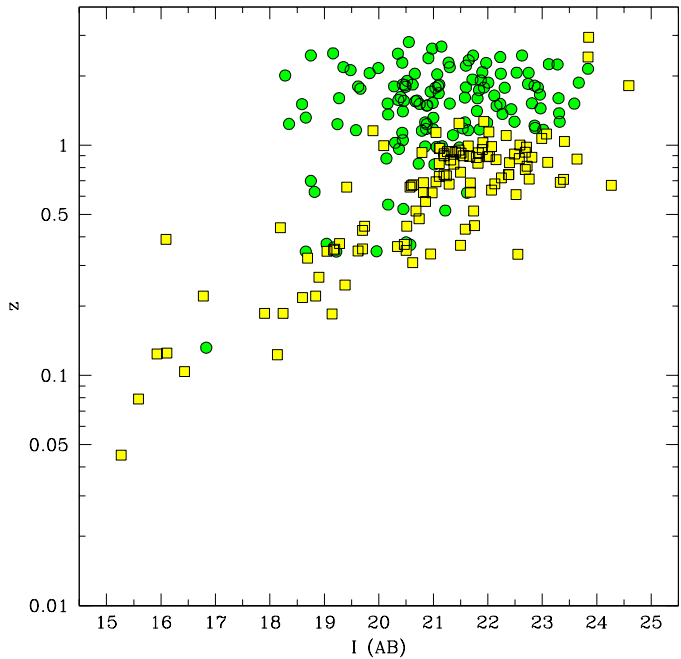}}
\caption{Redshift versus I-band magnitude for sources
  spectroscopically classified as BL AGN (green circles) and NOT BL AGN
  (yellow squares).}
\label{imagz}
\end{figure}
%\end{center}
%%%%%%%%%%%%%%%%%%%%%%%%%%%%%%%%%%%%%%%%%%%%%%%%%%%%%%%%%
%%%%%%%%%%%%%%%%%%%%%%%%%%%%%%%%%%%%%%%%%%%%%%%%
%\begin{center}
\begin{figure}
\centerline{\includegraphics[width=9.5cm]{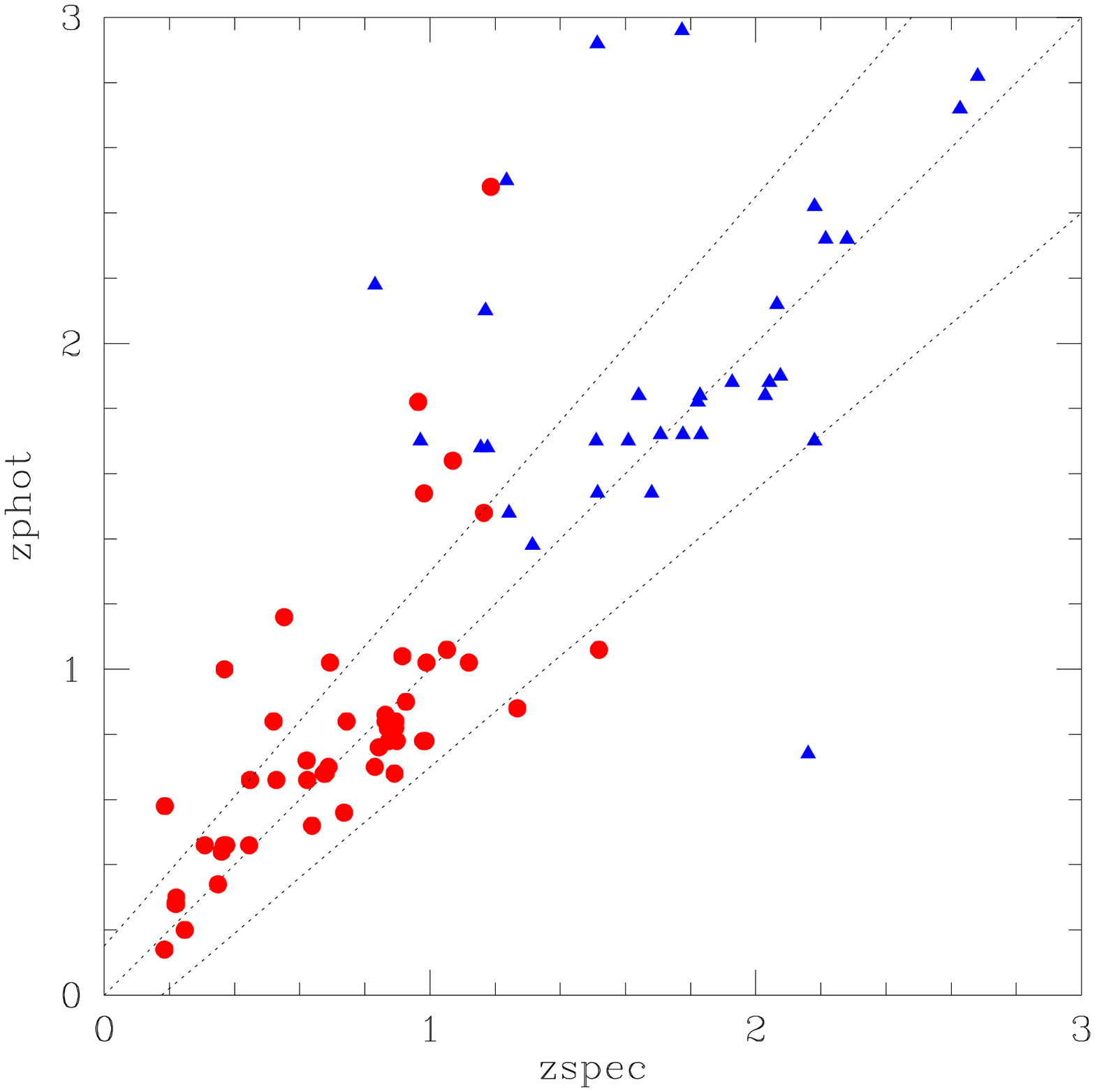}}
\caption{Photometric redshifts versus spectroscopic redshifts for the
  sources with secure redshifts in the 12F sample. Red circles mark extended
  sources, blue triangles mark point-like objects, on the basis of the ACS
  data. The dotted lines limit the locus for which $|z_{spec}-z_{phot}|<0.15\times(1+z_{\rm spec})$}.
\label{zphot}
\end{figure}
%\end{center}
%%%%%%%%%%%%%%%%%%%%%%%%%%%%%%%%%%%%%%%%%%%%%%%% 

\subsection{Results}
\pn
We computed the likelihood ratio for all the optical sources 
within 7\arcsec\ of the 674 X-ray centroids (a total of 2158 sources), 
finding 730 optical sources with LR$>$LR$_{\rm th}$. The expected number of true
identifications among these objects, computed by summing the reliability 
of each of them, is $\sim$524 ($\sim$78\% of the total number of X-ray
sources). The number of X-ray sources which have at least one optical source 
with LR$>$LR$_{\rm th}$ is 587 (87\% of the sample). The remaining 87 X-ray sources 
either are associated with LR$<$LR$_{\rm th}$ counterparts (74) or do not have any 
optical counterpart in the catalog within 7\arcsec\ from the X-ray centroids (13).   
We have therefore visually checked these 87 unidentified sources. 
In 24 cases we found a clear, relatively bright optical source 
within the X-ray error-circle in the I-band image.  These objects were missing in
  the input CFHT catalog used to compute the likelihood ratio (mainly because
  close to bright objects and/or in CCD masked regions) or they were 
  present but incorrectly associated with a much fainter magnitude. 
The I--band magnitudes of these 24 sources were therefore 
retrieved from the Subaru multicolor catalog, and added to our list of
possible counterparts.
It is important to note that this problem affects only a minority of the
X--ray source counterparts ($\sim$3.5\%, most of them close to bright stars
and/or defects in the optical band images) and it can be easily solved with
the adopted procedure.  \\
We then ran the likelihood analysis again, this time using as input catalog the K-band
(KPNO/CTIO) data extracted from the multicolor catalog (see
\citealt{capak}). The main advantage of using also this near-infrared catalog is
due to the fact that the X-ray to near-infrared correlation for AGN is much tighter than the one
in the optical bands \citep{mainieri02,brusa05}. Moreover, although the
currently available K--band data are shallower than the optical ones, their
use is potentially important to find the reddest optical counterparts, which
are a not negligible fraction of the identifications of faint X--ray sources
(see, e.g., \citealt{alex01}). 

Indeed, using this catalog  and adopting the same value of LR$_{\rm
    th}$=0.4, we found likely counterparts for 46 of the 87 unidentified
sources, that include all the 24 discussed above and 22 additional red, 
faint sources with unambiguous identification in the K--band (see also \citealt{mignoli}). \\
Summarizing, on the basis of the likelihood analysis applied to the I-band
and K--band catalogs, we have possible counterparts for 633 out of 674 X-ray sources
(654 out of 695 when the 21 sources brighter than I=16 are included).
We have further tentatively divided the 654 proposed identifications in
``reliable'' and ``ambiguous''. In the first class we include the 21 sources
with I$_{\rm AB}$$<$16, the 559 sources for which either there is only one object with
LR$>LR_{\rm th}$ or the ratio between the highest and the second
highest LR value in the I or K--band is greater than 3, and the 46 sources
identified only in the K--band. The total number of objects in this class is
therefore 626. The combined use of optical and NIR catalogs for the identification
process led to a number of "reliable" counterparts larger than that inferred 
from the use of the optical catalog only (524). However, we note that
$\sim5$\% of these reliable counterparts can still be misidentifications.
The other 28 sources comprise all the sources for which the ratio of LR values
in the I-- or K--band of the possible counterparts is smaller than 3, and they are
considered as ambiguous. In the following we tentatively identify these 28
X--ray sources with the optical counterparts with the highest LR value.\\
Table~1 gives a summary of the identifications, while Figure~2 shows a few
examples of finding charts for sources representative of the different types
discussed above.\\

\pn
For the remaining 41 sources, the possible
counterparts have on average faint or very faint optical magnitudes,
none of them has LR$>$ LR$_{\rm th}$, and most of them lie
at $\Delta(X-O)>$2$''$. For a fraction of them the true counterpart of
  the X--ray source may indeed be very faint and/or undetected in the optical
  bands (see e.g. \citealt{kok04}).  
Another possibility is that the X--ray emission originates from a group 
of galaxies at high-redshift with optically faint members: in this case, the X--ray extension of 
the putative group emission may be comparable to or smaller than the \xmm\ PSF 
and the source can have been classified as point-like.
Finally, a small fraction of these objects can be 
spurious X--ray sources, as suggested by the fact that the percentage of
sources with low X--ray detection likelihood (e.g. DETML $<9$; see
\citealt{papII} for a definition  of DETML) is significantly higher for these
unidentified sources ($\sim 30$\%) than for the optically identified X--ray
sources ($\sim 10$\%).
A detection with \chandra\, with significantly smaller
error--circles, would definitively discriminate among these different 
possibilities.

\subsection{Statistical properties}
\pn
The distribution of the X-ray to optical distances 
and of the I-band magnitudes for the X--ray sources identified as
described in Sect.~2.3 are shown in Fig.~3 (solid histograms). 
For the X-ray sources with more than one optical counterpart with
LR$>$LR$_{\rm th}$, the X-ray to optical distance and the I band magnitude of
the optical object with the highest LR is plotted. In the lower panel the
dashed histogram shows the lower limits to the I$_{\rm AB}$ magnitudes,
corresponding to the brightest objects  in the error circles, when the 41 X-ray
sources without an optical identification are also added. \\
About 90\% of the reliable optical counterparts have an X--ray to
optical separation ($\Delta(X-O)$) smaller than $3''$ (see inset in the upper
panel of Fig.~3), which is consistent with or even better than the $\sim 80$\%
within 3$''$ found in XMM--{\it Newton} data of comparable depth (see
e.g. \citealt{f03,dellaceca,l05}). The improvement is likely given by
  the combination of the accurate astrometry of the positions of the X-ray
  sources (tested with simulations in \citealt{papII}), and the full
  exploitation of the multiwavelength catalog (optical + K) for the
  identification process (that makes "reliable" a larger number of
  sources). In addition, because of the geometry of the final mosaic, most of
  the X--ray sources are closer than 5' to the center of at least one
  pointing, so that the corresponding PSF is not significantly
  deteriorated.
The majority of the X-ray sources ($\sim$73\%) have counterparts with I$_{\rm  AB}$ magnitudes in the range
$20<I_{\rm AB}<$24, with two tails at fainter ($\sim$12\%) and brighter
($\sim$15\%) magnitudes. The median magnitude of optical counterparts 
is I$_{\rm AB}=$21.65.\\
These distributions do not change significantly if we make a different choice
for the optical counterpart of the 28 ``ambiguous'' sources, i.e. the second
most likely optical counterpart is considered. Therefore, from a
statistical  point of view the properties of these 654 X-ray
sources (see Table 1) can be considered representative of the overall X-ray
point source population, at the sampled X--ray fluxes.

%%%%%%%%%%%%%%%%%%%%%%%%%%%%%%%%%%%%%%%%%%%%%%%%
%\begin{center}
\begin{figure*}
\centerline{\includegraphics[width=16cm]{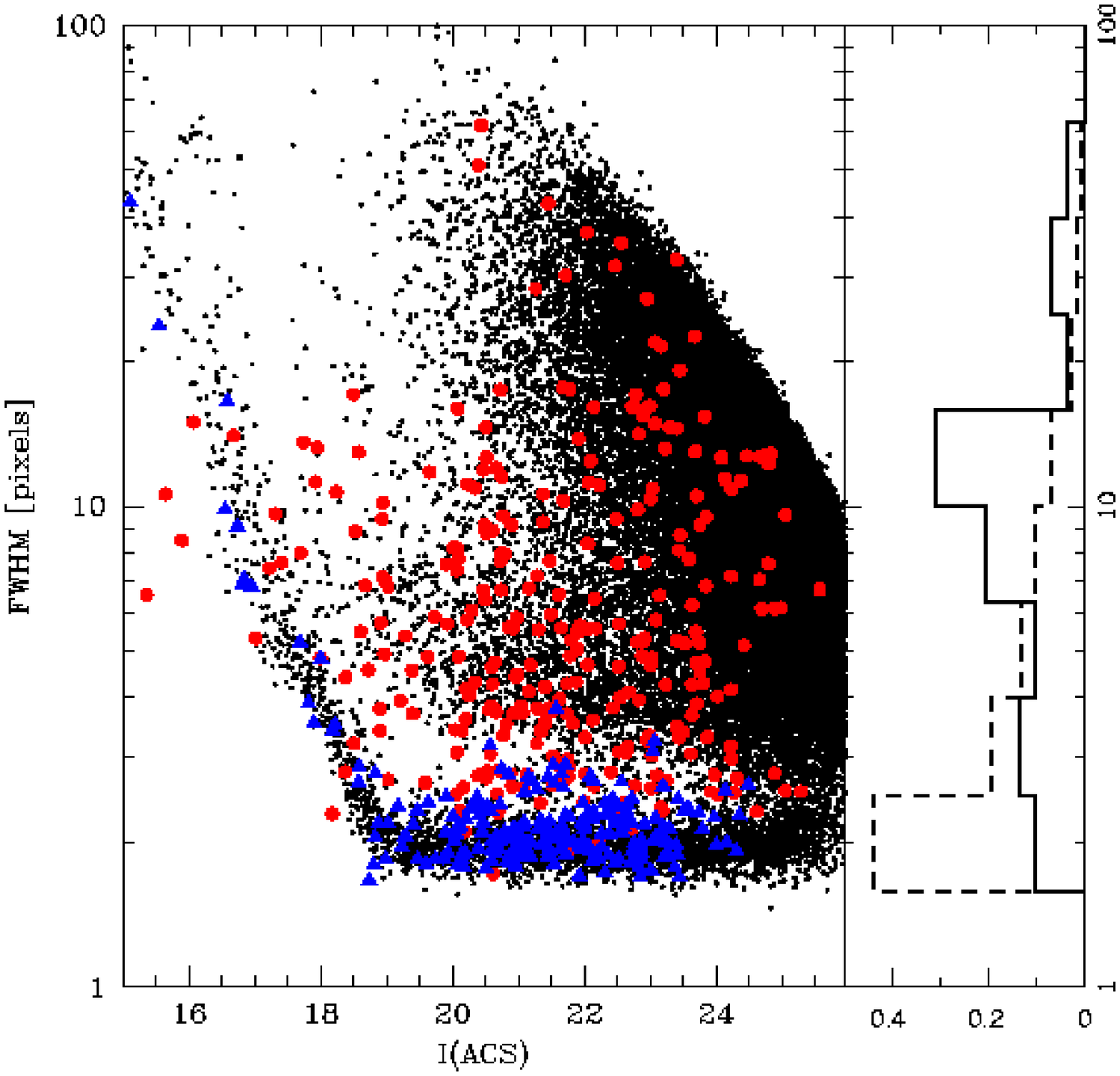}}
\caption{Full Width Half Maximum (FWHM) vs. ACS I-band magnitude for
  all the sources (small black symbols) in the ACS photometric catalog. 
 Overplotted as blue triangles (red circles) are the X-ray sources classified as
 point-like (extended) from ACS. 
 The  histograms on the right part of the figure show the distribution of FWHM
 for the sources detected only in the soft band (dashed histogram) and for the 
  sources detected only in the hard band (solid histogram). Both histograms
  are normalized to the total number of sources in each class.}
\label{fwhm}
\end{figure*}
%\end{center}
%%%%%%%%%%%%%%%%%%%%%%%%%%%%%%%%%%%%%%%%%%%%%%%%

\subsection{Redshift distribution}
Spectroscopic redshifts for the proposed counterparts are available from the
Magellan/IMACS observation campaign (193 objects, \citealt{impey,trump}) and  
the first run of zCOSMOS observations (25 objects, \citealt{lilly};
J.P. Kneib, private communication) and/or were already
present in the SDSS survey catalog (48 objects,
\citealt{sdss,kauf03}\footnote{These sources have been retrieved from the NED,
NASA Extragalactic Database}), for a total of 248 independent redshifts (five
of them of galactic stars). 
Spectroscopic information and classification is therefore available for a
substantial fraction of our optical/infrared counterparts (248/654,
$\sim$38\%). 
The X--ray to optical distances and the I--band magnitude distributions 
of the subsample with spectroscopic redshift are shown as shaded histograms in
both panels of Fig.~3. The spectroscopic completeness is $\sim$50\% at 
I$_{\rm AB}\leq $22 and drops to $\sim$35\% at 22$<$ I$_{\rm AB}\leq $23 and
to $\sim$18\% at 23$<$ I$_{\rm AB}\leq $24.\\ 
While a more detailed analysis of the optical spectra will be presented in
future papers, based also on more data that are rapidly accumulating from the
on-going projects, for the purposes of the present paper we divided the
extragalactic sources (243 objects) in only two classes, following
\citet{f03}: 
objects with broad optical emission lines (BL AGN, 126 in total) and sources
without a clear, broad emission signature in the optical band, but with narrow
emission or absorption  lines (NOT BL AGN, 117). The latter class is therefore
a mixed bag of different types of objects, and comprises mainly narrow-lined
AGN, low-luminosity AGN, starbursts and normal galaxies. 
Figure~4 shows the redshift distribution for all the extragalactic sources
in the spectroscopic sample (solid empty histogram) along with the
redshift distribution of the sources spectroscopically classified as BL AGN
(filled histogram). At low redshift (z$<$1) the spectroscopic 
identifications are dominated by objects  associated with NOT BL AGN,
while at high redshifts (z$>$1.5) almost exclusively BL AGN are detected. 
As already noted by \citet{sexsi} and \citet{cocchia}, the
small number of narrow line objects at z$>$1.5 is most likely due, at
least in part, to selection effects:
high-redshift narrow line AGN are on average fainter in the I-band than BL AGN
and a significant fraction of them is expected to be 
below the optical magnitude limit for spectroscopic follow-up. 
This is shown in Fig.~5, where the spectroscopic redshifts are plotted versus
the I--band magnitudes of the optical counterparts, for both BL AGN (green
circles) and  NOT BL AGN (yellow squares): while BL AGN show a large spread in
I magnitudes over a large redshift range, the distribution of NOT BL AGN in
this plane is much narrower and shows a rapid rise in I magnitude
with redshift (see also Fig. 14 in \citealt{trump} paper).
% hits the current spectroscopic limit (I$_{AB} \sim 23.5-24$) at z$\sim 1.5$.
Conversely, the small number of BL AGN at low-z can be ascribed to several factors: 
1) spectroscopic misclassification: given the wavelength range
($\sim$5400-9000 \AA) of both the IMACS and the VIMOS zCOSMOS data, at low  
redshift (z $\lsimeq$1) broad emission lines are not well sampled (e.g. H$\alpha$
is not visible at z$\gsimeq 0.4$, H$\beta$ is not visible at z$\gsimeq
0.8$ and MgII enters the spectral window at z$\sim$1), while narrow lines
(e.g. [OII]) and/or absorption features (e.g. Balmer break)
can be more easily detected;    
2) especially in the case of moderately weak Seyferts, the relative
contribution of the host galaxy and the AGN can be such that the broad lines are
diluted in the stellar light and can be revealed only in high 
S/N spectra \citep{moran,severgnini}.\\
\pn
To obtain a complete redshift distribution for the X--ray
sources, photometric redshifts should be computed for the sources without
spectroscopic data. As a training sample, we used the code described in
\citet{bender} and tested it on the sample of objects with secure
spectroscopic redshifts. 
For objects classified  as ``extended'' in the ACS images (see Section
  3.1),
we applied the same 
semi--empirical templates of non--active galaxies that have been used to compute
the photometric redshift in the Fornax Deep Fields (see \citealt{gabasch} and
references therein) and GOODS-South  field  \citep{salvato}; in  both applications, a $\sigma$($\Delta z/(1+z)$) $\sim$0.05
has been obtained. For objects that have been classified as point-like
sources, only AGN templates have been adopted (see also \citealt{mainieri05}
and reference therein).  
In Fig.~6 our best attempt so far is shown.  Red circles are  optically extended  X-ray
selected  sources, while blue triangles represent point-like objects.
A relatively high fraction of both extended ($\sim$73\%) and point-like ($\sim$66\%)
sources have $|z_{spec}-z_{phot}|<0.15\times(1+z_{\rm spec})$. 

%%%%%%%%%%%%%%%%%%%%%%%%%%%%%%%%%%%%%%%%%%%%%%%%
%\begin{center}
\begin{figure*}
\centerline{\includegraphics[width=16cm]{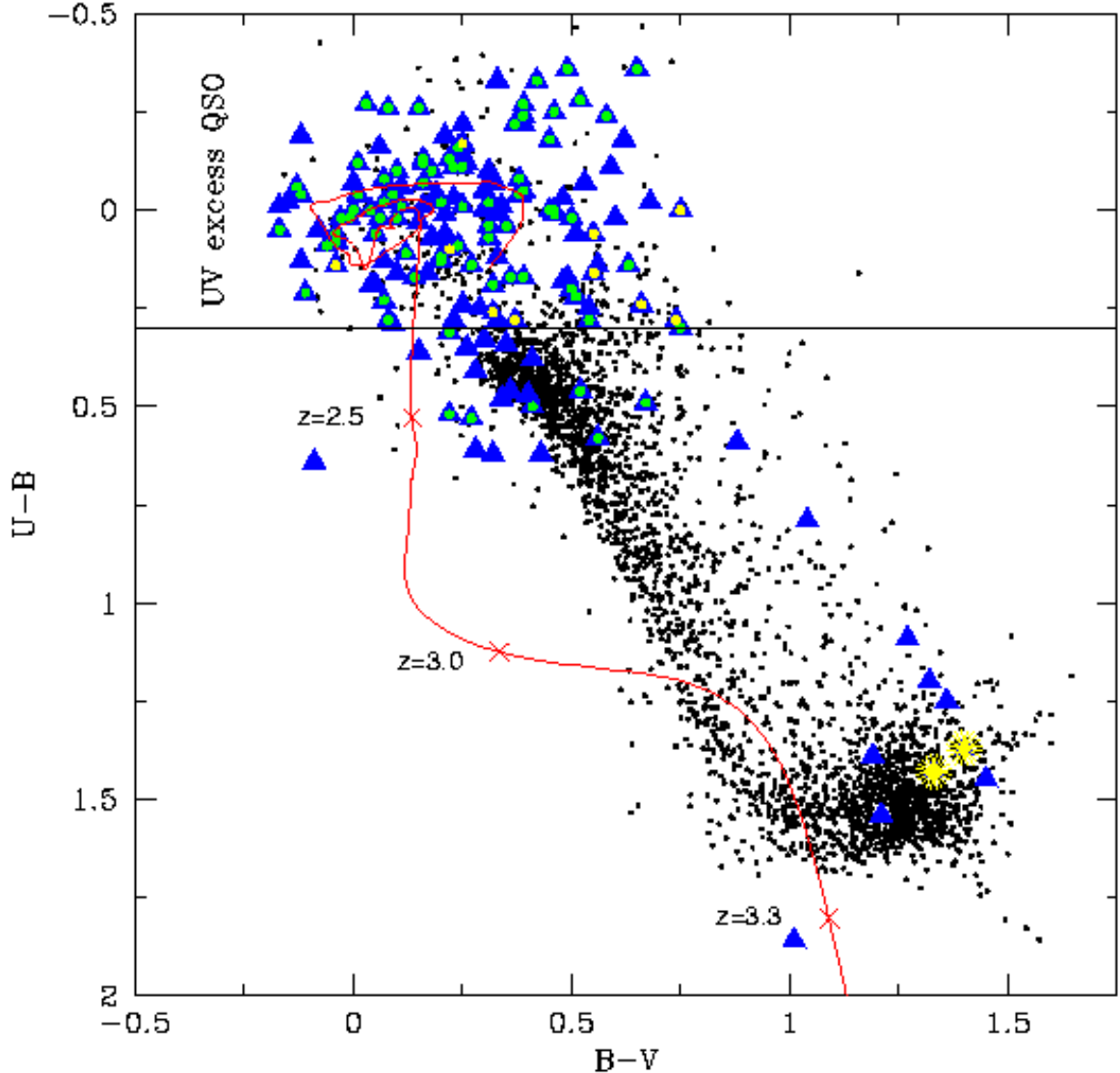}}
\caption{
U-B vs. B-V diagram for all the fields objects with B$>19$ (small
black points) classified as point-like in the full ACS catalog and with an
error in all the three optical bands smaller than $<0.15$ \citep{capak}.
The locus occupied by stars is clearly defined in this diagram,
with the two densely populated regions in the blue (upper left of the black points
envelope; hot subdwarf stars) and red (lower right; dwarf M stars) parts of the
sequence. The horizontal solid line marks the color cut for the selection of
ultraviolet excess objects. 
Overplotted as blue triangles are the counterparts of X--ray sources classified as point-like from ACS,
which satisfy the same selection in the photometric
errors. Spectroscopically confirmed BL AGN and NOT BL AGN are also indicated
(green and yellow symbols overplotted on the blue ones,
respectively). 
Finally, the expected tracks of quasars from redshift 0 to
$\sim$3.5 in this color-color diagram is also reported (see text for 
further details).}
\label{ubbv}
\end{figure*}
%\end{center}
%%%%%%%%%%%%%%%%%%%%%%%%%%%%%%%%%%%%%%%%%%%%%%%%

\section{Multiwavelength properties}
\subsection{ACS morphology: stellar vs. extended objects}
\pn
\pn
The catalog of the primary optical counterparts of the X--ray sources was then
cross-correlated with the June 2005 version of the ACS catalog
\citep{leauthaud,kok06}, from the first HST cycle (Cycle 12), in order
to gather some preliminary information on the morphological classification of
the X-ray sources.  Since the available ACS catalog does not cover the entire
XMM-COSMOS area analyzed here, we could use ACS information only for 524
optical sources, out of the total of 654. For these sources, we retrieved from
the catalog the measure of the Full Width Half Maximum (FWHM, in image 
pixels), the ACS I--band magnitude and a parameter that defines the
morphological classification (stellar or extended) on the basis of the 
analysis of the available data. Following \citet{leauthaud}, objects were
divided in ``stellar/point-like'' (hereinafter: point-like) and  ``optically
extended''' (hereinafter: extended) on the basis of their position in the
plane defined by the peak surface brightness above the background level and
the total magnitude (see Fig.~6 in \citealt{leauthaud}). 
The FWHM versus the I-band magnitude is shown in Fig.~7; 
in this figure, the blue triangles indicate sources classified as
``point-like'', and red circles sources classified as ``extended''. 
More than 50\% of the primary IDs have 
stellar (or almost stellar; FWHM $<$ 3 pixels) profile on ACS data. 
This is particularly true for the very soft sources (i.e. detected only in the
soft band, see dashed histogram in Fig.~7). 
The situation is completely different for the really hard sources 
(i.e. no detection in the soft band, solid histogram in Fig.~7), for which 
most of the counterparts ($\sim$80\%) are associated with extended 
sources. \\
A subsample of 214 objects (108 BL AGN, 101 NOT BL AGN, 5 stars) has
both spectroscopic information {\it and} ACS classification. In agreement with the
expectations, we find a good correspondence between ACS and spectroscopic
classification: the majority of BL AGN are classified as
point-like by ACS (86/108, $\sim$80\%), while the majority of NOT BL AGN are
classified as extended by ACS (85/101, $\sim$85\%). A total of 
38 sources show a ``mismatch'' between the morphological and spectroscopic
classification and we will come back to these sources later in the paper.

%%%%%%%%%%%%%%%%%%%%%%%%%%%%%%%%%%%%%%%%%%%%%%%%
%\begin{center}
\begin{figure*}
\centerline{\includegraphics[width=16cm]{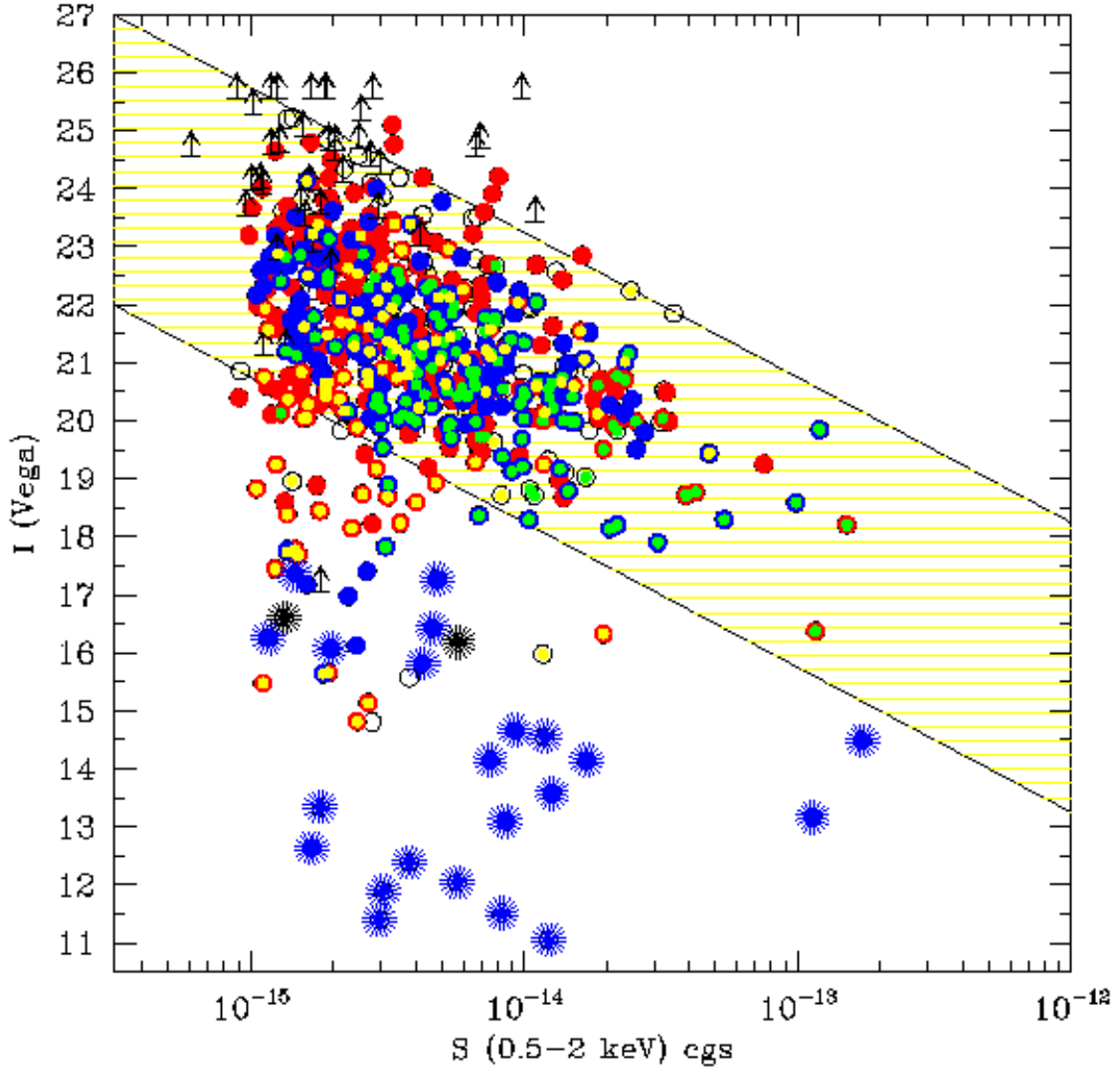}}
\caption{The I-band magnitude (Vega system) vs. the soft (0.5-2 keV) flux
  for all the primary counterparts in the sample (all symbols), and the subsamples
  of objects classified as point-like (blue circles) and extended (red circles)
  from ACS morphology. The shaded area represents the region typically occupied by known AGN
  (e.g. quasars and Seyferts) along the correlation log$(X/O)=0\pm
  1$. Spectroscopically confirmed BL AGN and NOT BL AGN are also indicated
  (green and yellow symbols, respectively). Asterisks mark the objects
  tentatively identified with stars (see Sect. 3.2 and 3.3). Lower limits to
  the I band magnitude correspond to the magnitude of the brightest objects in
  the error-circles of the unidentified X--ray sources. }
\label{Isx}
\end{figure*}
%\end{center}
%%%%%%%%%%%%%%%%%%%%%%%%%%%%%%%%%%%%%%%%%%%%%%%%

%%%%%%%%%%%%%%%%%%%%%%%%%%%%%%%%%%%%%%%%%%%%%%%%
%\begin{center}
\begin{figure*}
\centerline{\includegraphics[width=16cm]{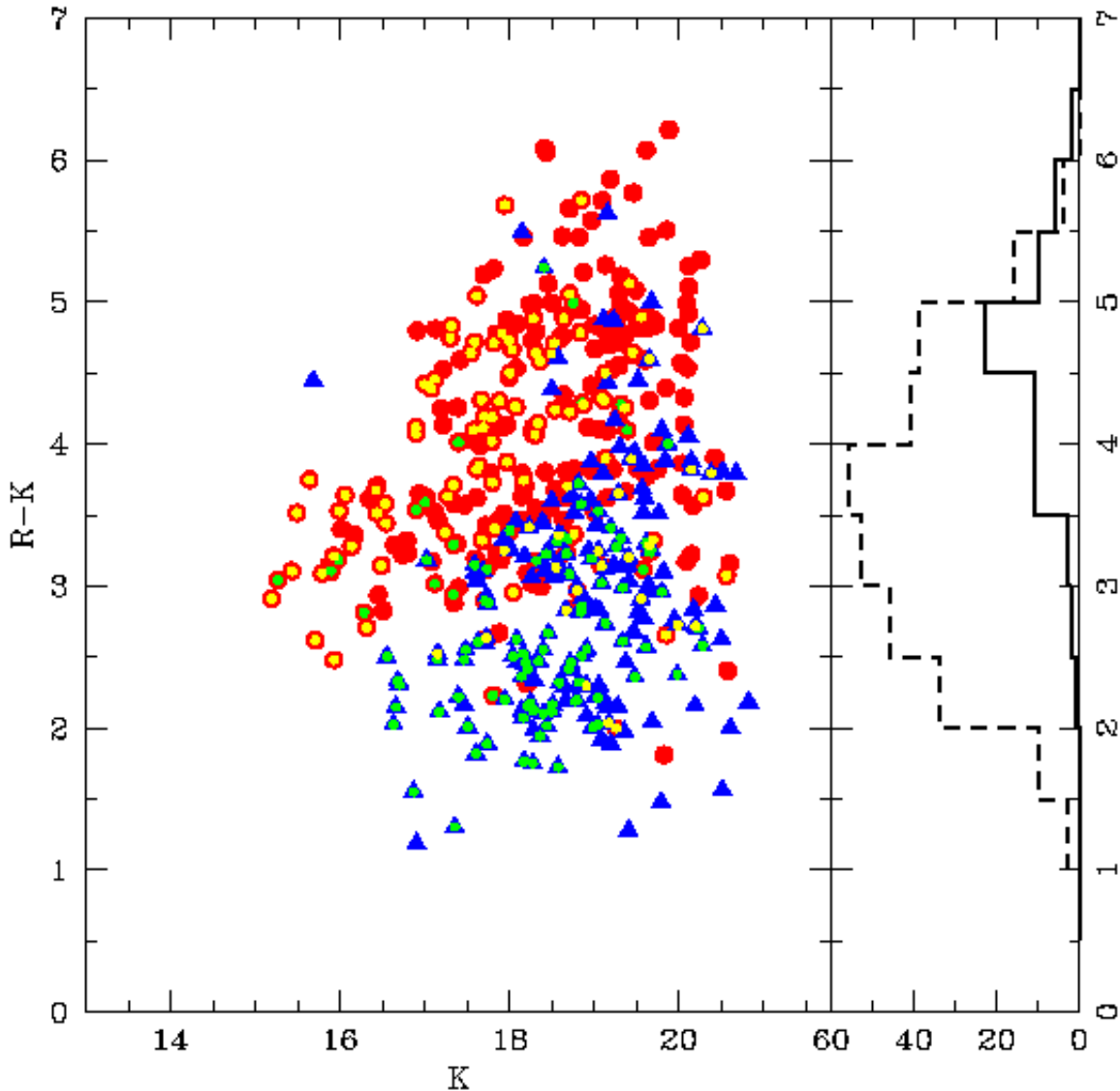}}
\caption{R-K color vs. K band magnitude (Vega system) for the X-ray sources
  counterparts with ACS information. Symbols as in Fig.~9. Sources
  spectroscopically identified with BL AGN (green) and NOT BL AGN 
  (yellow) are also highlighted. The histogram on the right part of
  the figure shows the distribution of R-K colors for the sources with
  HR$>-0.3$ (i.e. candidate X--ray obscured AGN, solid histogram) and the sources
  detected only in the soft band (HR=$-1$, dashed histogram).} 
\label{rkk}
\end{figure*}
%\end{center}
%%%%%%%%%%%%%%%%%%%%%%%%%%%%%%%%%%%%%%%%%%%%%%%%

\subsection{Optical color-color diagrams for point-like objects}
\pn
Figure~8 shows the U-B vs. B-V diagram for all field objects (small
black points) with B$>$19 classified as point-like in the full ACS catalog and with an
error in all the three optical bands smaller than 0.15 magnitude \citep{capak}.
The locus occupied by stars is very well defined in this diagram, with the two
densely populated regions in the blue (upper left of the black points
envelope; hot subdwarf stars) and red (lower right; dwarf M stars) parts of the
sequence. Most of the points at U-B$<$0.3 (UVX objects) are expected to be  
AGN at z$\lsimeq$2.3, while the objects on the right of the stellar sequence
are likely to be compact galaxies classified as point-like. Overplotted as blue
triangles are the X--ray sources classified as 
point-like from ACS which satisfy the same selection in the photometric
errors. Spectroscopically confirmed BL AGN and NOT BL AGN are also indicated
(following Fig.~5, green and yellow symbols overplotted on the blue ones, 
respectively). Finally, the expected track of quasars from 
redshift 0 to 3.5 in this color-color diagram is also reported, computed using
the SDSS AGN template from (\citealt{buda01}; see also \citealt{capak}
for further details).\\
It is quite reassuring that the majority ($\gsimeq 80$\%) of X-ray selected point-like quasars
occupy the classical QSO locus and would have been selected as outliers from
the stellar locus in this color-color plot. 
A few (5 to 8) sources are within or close to the locus of ``red'' stars. 
So far we have spectroscopic data only for two of these objects and both of
them are stars. All these sources are detected only in the
soft X-ray band and show a low ($<$0.1) X-ray to optical flux ratio (X/O)
\footnote{The I--band flux is computed by
converting I magnitudes into monochromatic  fluxes and then multiplying 
them by the width of the I filter \citep{zombeck}. }
and are therefore likely to be stars.
A similar number (8 to 10) of X-ray sources are within or close to the locus of
blue stars.  
The spectroscopic identifications available for these objects include three BL AGN
at z$\geq1.4$; a fourth object is likely to be a star on
the basis of the bright I band magnitude and low X/O ratio. \\

\subsection{X--ray to optical flux ratio}
\pn
Another independent check on the agreement between optical
identification, morphological analysis and spectroscopic breakdown 
is shown in Fig.~9, where the I-band magnitude (Vega system) is plotted versus the
soft X-ray flux for all the X-ray sources (black empty symbols), ACS-point-like
objects (blue symbols) and ACS-extended sources (red symbols). \\
As discussed in Sect~2.1, the sources brighter than I$_{\rm AB}$=16 are
identified ``a priori'': all of them have been visually inspected and the
majority of them turned out to be associated with bright stars and have
point-like morphology from ACS. A sample of 8 additional stars has been
tentatively identified on the basis of the information from U$-$B vs. B$-$V
diagram, the bright optical magnitudes and the low X--ray to optical flux
ratio. All the sources identified with stars (a total of 25) are plotted as
asterisks in Fig.~9 and they constitute $\sim$4\% of the soft X--ray
selected sample. This value is similar to that reported in the ROSAT Deep
Survey \citep{l01} and in the ChaMP survey \citep{green04}, from the optical
identification of large samples (100--300) of soft X--ray selected sources at
limiting fluxes similar to that of COSMOS.  \\

\pn
If objects associated with stars are removed, point-like sources occupy
preferentially the locus in this plane with X/O in the range 0.1$\div$10, where most of
the broad line quasars from previous optical and soft X--ray selected surveys
lie (hatched region in Fig.~9). 
Also most of the extended sources are within the same range of X/O ratio, but
with more significant tails toward both low and high values of X/O.
For most of these sources, thanks to the superb ACS resolution, it will be possible to resolve the
nucleus and the host galaxy. 
Optically bright (i.e. X/O$<$0.1) extended sources are
preferentially identified with nearby galaxies, in which the optical
luminosity is mainly due to the integrated stellar light; for the majority of
them, the high-level of observed X--ray flux suggests that some activity is
taking place in their nuclei (see, e.g. \citealt{xbong}). 

\subsection{Optical to near-infrared colors} 
\pn
Figure~10 shows the R$-$K vs. K--band magnitude (Vega magnitudes) for the subsample of objects with
ACS morphological information (excluding spectroscopically confirmed stars). In
addition, sources spectroscopically identified with BL AGN are marked with
green symbols, while sources identified with NOT BL AGN are marked by
yellow symbols. A significant difference in the R$-$K distributions for
point-like and extended sources is present : while the widths of the two
distributions are similar ($\sigma\sim$0.8 for both of them), extended sources
are significantly redder ($\langle R-K\rangle$=4.05$\pm0.05$) 
than point-like sources 
($\langle R-K\rangle$=2.91$\pm0.06$). 
When the spectroscopic information is also considered, objects with red $R-K$
colors are preferentially associated with NOT BL AGN (yellow circles), while
blue objects are preferentially associated with BL AGN (green circles).\\
\pn
We have then investigated the distribution of the $R-K$ colors as a function
of the X--ray hardness ratio (HR), a widely used tool to study the
  general X-ray spectral properties of X-ray sources when the number of
  counts is inadequate to perform a spectral fit. \citet{papIV} 
  (paper IV of this series) have shown that 99\% of the sources 
 with HR$>-0.3$ in their subsample can be fit by a $\Gamma$=1.8-2 power-law
 continuum absorbed by a column density (N$_H$) larger than
 $10^{22}$ cm$^{-2}$. 
 Conversely, sources detected only in the soft band (i.e. HR=-1) are 
 most likely unobscured.  \\
The hardest sources (HR$>-0.3$, solid histogram in the right part of
Fig.~10) are mostly associated with red and very red objects (R-K$>$4). This
indicates an excellent consistency between optical obscuration of the nucleus
as inferred from ACS (extended morphology), optical to near infrared colors,
and the presence of X--ray obscuration as inferred from the hardness ratio
(see also \citealt{alex01,g02,brusa05,papIV}). Conversely, sources
detected only in the soft band (i.e. HR=$-1$, dashed histogram) have
preferentially blue R-K colors ($>60\%$ of the sources have $R-K<4$), typical
of those of optically selected, unobscured quasars \citep{bar01}. 
It is interesting to note, however, that the
observed correspondence between hard (soft) X--ray colors and red (blue)
optical to near infrared colors is not a one-to-one correlation (see also,
e.g., \citealt{georg04,brusa05}). In fact, as shown for example by the tail
toward high R-K colors of the dashed histogram in Fig.~10, a non-negligible
fraction of red objects is associated with very soft X--ray sources. \\

%%%%%%%%%%%%%%%%%%%%%%%%%%%%%%%%%%%%%%%%%%%%%%%%
%\begin{center}
\begin{figure}
\centerline{\includegraphics[width=9.5cm]{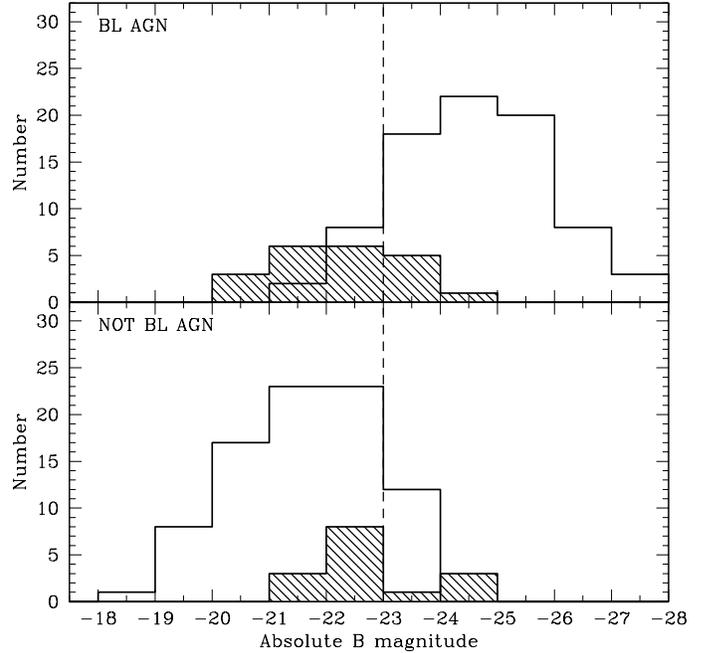}}
\caption{
  Distribution of the absolute B--band magnitude for BL AGN (solid histogram, upper panel) and
  NOT BL AGN (solid histogram, lower panel) in the subsample of objects with ACS information
  and spectroscopic classification. 
  In both panels the shaded histograms give the contribution of outliers in
  each class, i.e. ``extended'' BL AGN and ``point--like'' NOT BL AGN, respectively.
  The dashed line at M$_{\rm B}$=-23 marks the ``classic'' separation between
  Seyfert and Quasars.} 
\label{mbhisto}
\end{figure}
%\end{center}
%%%%%%%%%%%%%%%%%%%%%%%%%%%%%%%%%%%%%%%%%%%%%%%%

%%%%%%%%%%%%%%%%%%%%%%%%%%%%%%%%%%%%%%%%%%%%%%%%
%\begin{center}
\begin{figure}
\centerline{\includegraphics[width=9.5cm]{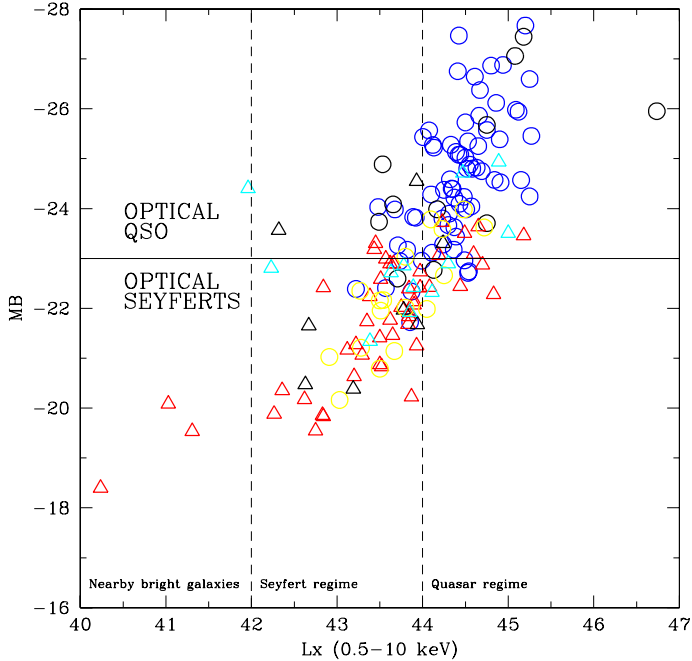}}
\caption{Absolute rest frame B-band magnitude vs. the logarithm of the full
  (0.5-10 keV) X--ray luminosity (erg s$^{-1}$). 
  Symbols are as follows: Blue circles = spectroscopically
  confirmed BL AGN classified as {\it point-like} from ACS; 
  Yellow circles = spectroscopically confirmed BL AGN classified as {\it
    extended} from ACS; 
  Cyan triangles = spectroscopically confirmed NOT BL AGN
  classified as {\it point-like} from ACS;    
  Red triangles = spectroscopically confirmed NOT BL AGN
  classified as {\it extended} from ACS. The horizontal solid line marks the
  ``classic'' optical transition between Seyfert and Quasars.}
\label{lxmb}
\end{figure}
%\end{center}
%%%%%%%%%%%%%%%%%%%%%%%%%%%%%%%%%%%%%%%%%%%%%%%% 

\subsection{Outliers: mismatch between morphological and spectroscopic
  classification} 
\pn
As discussed in Sect.~3.1, there is a relatively good agreement (of the order
of 80\%) between the morphological and spectroscopic classifications for the
subsample of sources for which these informations are available.
However, there are 22 ($\sim 20$\%) BL AGN classified as ``extended'' in ACS
and 16  ($\sim 16$\%) NOT BL AGN classified as point-like by ACS. \\
The absolute magnitude is an obvious quantity to examine in order to try to
  understand why for these sources the morphological and spectroscopic
  classifications do not agree with each other.
Figure~11 shows the distribution of rest--frame, absolute B--band magnitude for the
209 extragalactic sources which have both morphological (ACS) and spectroscopic classification, separately for BL
AGN (upper panel) and NOT BL AGN (lower
panel). The black filled histograms show the observed distribution for the
``outliers'', i.e. ``extended'' BL AGN and ``point-like'' NOT BL AGN,
respectively. In order to minimize the uncertainties in the K-correction, the absolute B--band magnitudes have
been computed from the apparent magnitude in the optical filter closest to the
rest-frame B band at any given redshift. \\
All the ``extended'' BL AGN lie at redshift $<$1.5, while the redshift
distribution of the full BL AGN population peaks at z$\sim$1.5 (see Figg.~4
and 5). The fact that ACS reveals an extended component is likely to be due to
the fact that the nuclear emission in these objects is not dominant  with
respect to the host galaxy emission, at least in the observed ACS band.  
Indeed, the average M$_B$ of the extended BL AGN sources lies in the low-tail of the M$_B$
distribution of the entire BL AGN population and typically in the Seyfert
regime, taking M$_B=-23$ as the ``classical'' separation 
between Seyferts and QSO \citep{sg83}. \\ 
  The unresolved nature of ``point-like'' NOT BL AGN is more puzzling. 
  It can be due, at least in
  part, to the definition we adopted for the segregation of stellar objects:
  especially at high redshift, the ratio between the peak flux and the total
  magnitude cannot be such to define them "extended".   
  However, we note that
  most of them occupy the high-luminosity tail of the distribution of M$_B$ of
  the overall population of NOT BL AGN, and therefore can be classified as 
  high-luminosity, type 2 quasars from the available optical spectroscopy (see
  also \citealt{papIV}  for a more detailed analysis of the rest--frame X--ray
  properties). Moreover, they tend to be ``bluer'' (i.e. less absorbed) in
  the optical to near infrared colors (yellow symbols on blue points in
  Fig.~10), suggesting that we are seeing more directly the
  emitting nucleus. 
 
\pn
%The same trend is also present when X--ray luminosities are considered. 
Fig.~12 shows the absolute B magnitude as a function of the 0.5-10 keV 
luminosity (as computed in \citealt{papIV}) for the subsample of
sources with spectroscopic redshift and good X--ray counting statistics. 
Circles mark BL AGN (blue for the point-like and yellow for the extended
sources) while triangles mark NOT BL AGN (cyan for the point-like and red for
the extended sources).   
Sources with L$_X<10^{42}$ erg s$^{-1}$ are mostly associated with bright
(I$<$18), extended objects (red triangles) at z$<$0.2; two of them  are detected
only in the soft band. 
Among the two sources detected also at higher energies, there is source the
candidate Compton Thick AGN discussed in \citet{papI}.\\ 
The region at luminosities higher than 10$^{44}$ erg s$^{-1}$ is mainly populated by
point-like, BL AGN (blue circles). Only 8 sources classified as NOT BL
AGN lie in this part of the diagram: the high X--ray luminosity classifies
them as candidate Type 2 QSO \citep{papIV}. 
Conversely, the majority of the sources in the middle part of the X--ray
luminosity region are
associated with extended, NOT BL AGN (red triangles) and low redshifts BL
AGN (Seyfert 1 galaxies, yellow circles).  \\

\section{Discussion}
\pn
Comparing the optical color-color properties of AGNs in the COSMOS
field with those of field objects (see Fig.~8), we estimate that X--ray data 
with a flux limit of S$_{\rm 0.5-2 keV}\sim10^{-15}$ erg cm$^{-2}$ s$^{-1}$ recover at
least half (50\%) of the AGN candidates which would be selected on the basis of their ultraviolet excess
(U-B$<$0.3) at B$\lsimeq$23 (see \citealt{zamorani} for a similar estimate at
a somewhat brighter X--ray flux from {\it ROSAT} observations in the Marano
field). This fraction of 50\% has been obtained assuming that all objects with
ultraviolet excess are AGN. Since at these magnitudes blue stars and compact
galaxies may represent $\sim 30$\% of the samples of point--like objects with
ultraviolet excess \citep{migno_gz}, we conclude that the efficiency of present
X--ray data in detecting AGN with ultraviolet excess at B$\leq$23 is likely to
be as high as 70\%. We note that the large majority of the BL AGN
  selected via the UV excess method have MB$<-23$ (see Fig. 11) and are
  therefore classified as quasars, confirming previous results (see e.g
  \citealt{zamorani} and reference therein).   
We also note, however, that the photometric catalog we use has deeper limiting
magnitudes with respect to all the previous studies on optically selected
UV-excess QSO and therefore we start to
explore and pick up lower-luminosity (e.g. Seyfert-like) objects up to
z$\sim1.5$.  
On the other hand, a significant fraction ($\sim$40\%) of the X-ray selected
AGNs, especially those without broad lines in their optical spectra, would
have not been easily selected as AGN candidates on the basis of purely optical
criteria, either because of colors similar to those of normal
stars or field galaxies at z$\sim$1-3, or because of morphological
classification not consistent with that of point--like sources (see Fig.~10).  
These results highlight the need for a full multiwavelength coverage to
properly study and characterize the AGN population as a whole. \\
With the current spectroscopic coverage ($\sim$38\%)
the observed redshift distribution and spectroscopic classification 
of identified X--ray sources could be biased against low-redshift BL AGN
and high-redshift NL AGN (see Sect.~2.5 and \citealt{trump} for a more detailed
discussion on AGN selection effects in COSMOS; see also \citealt{treister}). 
However, according to the most recent modeling of the X--ray luminosity
function evolution (e.g. \citealt{ueda,clasxs2,has05,lafranca}), the paucity
of  low-redshift BL AGN and high--redshift obscured NL AGN may be at least
partly real and due to a luminosity dependent evolutionary behaviour of X--ray
selected sources. 
The space density of high luminosity (L$_{\rm X}>10^{44}$ erg s$^{-1}$) quasars
decreases steeply below z$\simeq$1, while the decrease of the space density
of lower luminosity objects in the same redshift range is much slower (see,
e.g., Fig.~9 in \citealt{f03}). 
Moreover, there are now increasing evidences that the relative ratio
  between 
obscured and unobscured AGN is a decreasing function of the X--ray luminosity
\citep{ueda,hasinger04,lafranca,akylas}. 
As a consequence, at low redshifts the X--ray source population is 
dominated by low luminosity, often obscured AGN, while at higher redshifts
luminous unobscured quasars take over. This is in qualitative agreement with
the observed change of the Type 2/ Type 1 ratio (R) in the present sample  
when the typical Seyferts (32/33, R$\sim$1 at L$_X\sim 10^{42}-10^{44}$ erg
s$^{-1}$) and Quasars (8/44, R$\sim$0.18 at L$_X>10^{44}$ erg s$^{-1}$) 
luminosities are considered (see Fig.~12).\\

At a limiting flux of 2$\times 10^{-15}$ erg cm$^{-2}$ s$^{-1}$ in the soft band, where
the sky coverage has decreased to 50\% of the total 1.3 deg$^2$ area covered by
the 12F pointings, no object at z$>$3 is present so far in
our current spectroscopic sample.  
On the one hand, the lack of high redshift sources could be due, at least
in part, to the still limited spectroscopic completeness, which is of the order of 50\%
at I$_{\rm AB}$$<$22 and about 25\% at fainter magnitudes, with no
spectroscopic redshifts for I$_{\rm AB}$$>$24. 
As a comparison, \citet{murray} have 14 z$>3$ quasars in their 9 deg$^2$
X-Bootes survey, at a limiting flux about one order of magnitude brighter
than our survey and with a similar spectroscopic completeness. By rescaling 
the area and the limiting fluxes, this would translate in $\sim 3$ z$>3$
quasars  
in the XMM-COSMOS 12F area, that is still consistent with zero being observed. 
On the other hand, the number of AGN at z$>$3 expected from population
synthesis models (see also \citealt{papI,gilli06}) is $\sim$30 and a random
sampling at $\sim$25\% completeness, to take into account the current
follow-up spectroscopy, would therefore predict about 7 quasars at
z$>$3, which are not observed. 
The forthcoming spectroscopic observations will increase the spectroscopic
completeness, allowing to verify if a sizable number of high redshift (z$>$3) quasars is 
present among the optically faintest counterparts and among the optically
unidentified X--ray sources, or if some of the model assumptions,  mainly
  based on extrapolations of the lower redshift (z$<3$) luminosity functions,
should be refined. \\

\section{Summary}
\pn
In this paper we presented the optical identifications 
for a large subsample ($\sim$700) of X--ray sources detected in the first 1.3
deg$^2$ of the XMM-COSMOS observations, down to a 0.5-2 keV limiting flux of
$\sim10^{-15}$ erg cm$^{-2}$ s$^{-1}$. 
The X-ray counterparts have been identified in optical (I--band) and near
infrared (K--band) catalogs, making use of the ``maximum likelihood''
technique. The combined use of the two different catalogs allowed us to test
the identification procedure and turned out to be extremely useful  
both for isolating input catalogs problems and for identifying optically faint
counterparts. Overall, 90\% (626) of the X--ray sources have been
unambiguously associated with optical or near--infrared counterparts. 
Twenty eight X--ray sources have 2 possible optical
counterparts with comparable likelihood of being the correct identification.
These sources have been classified as ``ambiguous'' and we tentatively
identify them with the optical counterpart with the highest LR value.
The sample of proposed identifications therefore comprises 654 objects, 
while for the remaining 41 X--ray sources it was not possible to assign a
candidate counterpart. We will use the forthcoming {\it Chandra} data (granted
in AO8) to definitively discriminate the ambiguous and false associations and
we also predict that a large fraction of these very faint, unidentified 
objects will be resolved with the inclusion of {\it Spitzer} (IRAC and MIPS)
data in the source identification. \\
We then cross-correlated our proposed optical counterparts with the
Subaru multi-color catalog \citep{capak}, the HST/ACS data \citep{leauthaud} 
and the first results from the massive spectroscopic campaigns in the COSMOS
field \citep{trump,lilly}. 
Our analysis reveals that for $\sim$80\% of the X--ray
sources counterparts with spectroscopic redshifts (a total of 248) there is a good
agreement between the spectroscopic classification, the morphological
parameters, and optical to near infrared colors: the large majority of
spectroscopically identified BL AGN have a point-like morphology on ACS data
(see Sect.~3.1), blue optical colors in color-color diagrams (see Sect.~3.2),
and an X--ray to optical flux ratio typical of optically selected quasars (see
Sect.~3.3). Conversely, NOT BL AGN are on average associated with extended
optical sources, have significantly redder optical to near infrared colors and
span a larger range of X--ray to optical flux ratios (see Sect.~3.3 and 3.4). 
When the X--ray information is also considered, we found that hard X--ray
sources are preferentially associated with extended sources and are ``reddish''
($\sim$60\% with R-K$>$4, see also \citealt{papIV}), while sources detected
only in the soft band are mostly associated with point-like objects and have
``blue'' optical to near--infrared colors. \\
Comparing the optical color-color properties of AGNs in the COSMOS
field with those of field objects (see Fig.~8), we estimate that X--ray data 
with a flux limit of S$_{\rm 0.5-2 keV}\sim10^{-15}$ erg cm$^{-2}$ s$^{-1}$
recover $\sim 70$\% of the AGN selected on the basis of their ultraviolet excess
at B$\lsimeq$23. This fraction will rise up to (90-100)\% at S$_{\rm 0.5-2
  keV}\sim5\times10^{-16}$ erg cm$^{-2}$ s$^{-1}$, the final depth of the 
XMM--COSMOS survey. \\
About 20\% of the sources show an apparent mismatch between the morphological
and spectroscopic classifications. Our analysis indicates that, at least for
BL AGN, the observed
differences are largely explained by the location of these objects in the
redshift--luminosity plane (see sect.~3.5). Our analysis also suggests a
change of the Type 2/Type 1 ratio as a function of the X--ray luminosity, in
qualitative agreement with the results from X--ray spectral analysis
\citep{papIV} and the most recent modeling of the X--ray luminosity function
evolution (Figg.~5 and~12). \\
Although the Magellan/IMACS and VIMOS/zCOSMOS spectroscopic campaigns will
continue to obtain redshifts for AGN in the COSMOS field, we expect that a
fraction of the order of 30\% of the sources will not have spectroscopic
redshifts, due to the faintness of their optical counterparts ($\sim 15$\%)
and the limitation of multi-slit spectroscopy due to mask efficiency \citep{impey}. 
We have tested a photometric code specifically designed to obtain redshifts for
X--ray selected sources on the subsample of sources with spectroscopic data,
and showed that it is possible to obtain a reasonable estimate of the
sources redshifts ($|z_{spec}-z_{phot}|<0.15\times(1+z_{\rm spec})$), for
$\sim$70\% of the sources. We
plan to use the same code to derive the redshifts for the faintest X--ray
counterparts. \\  
\pn
Summarizing, we were able to perform a combined multicolor, spectroscopic and
morphological analysis on a statistical and meaningful sample of X--ray
sources,  and this work represents one of the most comprehensive
multiwavelength studies to date of the sources responsible of most of the XRB 
(see also \citealt{sexsi}). 
This work is just the first phase of COSMOS AGN studies; the results presented
in this paper clearly show the need for a full multiwavelength coverage to
properly study and characterize the AGN population as a whole. 
As more data will become available,  the selection of COSMOS AGNs by all
available means - X-ray, UV,  optical, near-IR, and radio - will build up to
give the first bolometric selected  AGN sample, fulfilling the promise of many
years of multi-wavelength studies  of quasars.

\acknowledgments
This work is based on observations obtained with XMM-{\it Newton}, 
an ESA Science Mission with instruments
and contributions directly funded by ESA Member States and the
USA (NASA). In Germany, the XMM-{\it Newton} project is supported by the
Bundesministerium f\"ur Wirtschaft und Technologie/Deutsches Zentrum
f\"ur Luft- und Raumfahrt (BMWI/DLR, FKZ 50 OX 0001), the Max-Planck
Society and the Heidenhain-Stiftung. 
Part of this work was supported by the Deutsches Zentrum f\"ur Luft-- und
Raumfahrt, DLR project numbers 50 OR 0207 and 50 OR 0405. In Italy, the
XMM-COSMOS project is supported by ASI-INAF and MIUR under grants I/023/05/00
and Cofin-03-02-23. MB and GZ gratefully acknowledge useful discussion with 
G. Micela and S. Sciortino. 
We thank the anonymous referee for her/his useful comments on this manuscript. 
We gratefully acknowledge the contributions of the entire COSMOS
collaboration; more information on the COSMOS survey is available at 
\verb+http://www.astro.caltech.edu/~cosmos+. This research has made  
use of the NASA/IPAC Extragalactic Database (NED) and the SDSS spectral 
archive.

\clearpage

\end{document}